






  \font\twelverm=cmr10 scaled 1200       \font\twelvei=cmmi10 scaled 1200
  \font\twelvesy=cmsy10 scaled 1200      \font\twelveex=cmex10 scaled 1200
  \font\twelvebf=cmbx10 scaled 1200      \font\twelvesl=cmsl10 scaled 1200
  \font\twelvett=cmtt10 scaled 1200      \font\twelveit=cmti10 scaled 1200

  \font\twelvemib=cmmib10 scaled 1200
  \font\elevenmib=cmmib10 scaled 1095
  \font\tenmib=cmmib10
  \font\eightmib=cmmib10 scaled 800


\font\elevenrm=cmr10 scaled 1095    \font\eleveni=cmmi10 scaled 1095
\font\elevensy=cmsy10 scaled 1095


\font\seventeeni=cmmi10 scaled \magstep3

\font\seventeensy=cmsy10 scaled \magstep3

\font\seventeenmib=cmmib10 scaled \magstep3

\newfam\cpfam%


 \font\eightbf=cmbx8


\skewchar\eleveni='177   \skewchar\elevensy='60
\skewchar\elevenmib='177  \skewchar\seventeensy='60
\skewchar\seventeenmib='177
\skewchar\seventeeni='177

\newfam\mibfam%


  \skewchar\twelvei='177   \skewchar\twelvesy='60
  \skewchar\twelvemib='177
%
%
\def\twelvepoint{\normalbaselineskip=12.4pt
  \abovedisplayskip 12.4pt plus 3pt minus 9pt
  \belowdisplayskip 12.4pt plus 3pt minus 9pt
  \abovedisplayshortskip 0pt plus 3pt
  \belowdisplayshortskip 7.2pt plus 3pt minus 4pt
  \smallskipamount=3.6pt plus 1.2pt minus 1.2pt
  \medskipamount=7.2pt plus 2.4pt minus 2.4pt
  \bigskipamount=14.4pt plus 4.8pt minus 4.8pt
  \def\rm{\fam0\twelverm}          \def\it{\fam\itfam\twelveit}%
  \def\sl{\fam\slfam\twelvesl}     \def\bf{\fam\bffam\twelvebf}%
  \def\mit{\fam 1}                 \def\cal{\fam 2}%
  \def\tt{\twelvett}%
  \def\mib{\fam\mibfam\twelvemib}%

  \textfont0=\twelverm   \scriptfont0=\tenrm     \scriptscriptfont0=\sevenrm
  \textfont1=\twelvei    \scriptfont1=\teni      \scriptscriptfont1=\seveni
  \textfont2=\twelvesy   \scriptfont2=\tensy     \scriptscriptfont2=\sevensy
  \textfont3=\twelveex   \scriptfont3=\twelveex  \scriptscriptfont3=\twelveex
  \textfont\itfam=\twelveit
  \textfont\slfam=\twelvesl
  \textfont\bffam=\twelvebf \scriptfont\bffam=\tenbf
                             \scriptscriptfont\bffam=\eightbf
  \textfont\mibfam=\twelvemib       \scriptfont\mibfam=\tenmib
                               	      \scriptscriptfont\mibfam=\eightmib

  \def\xrm{\textfont0=\twelverm\scriptfont0=\tenrm
      \scriptscriptfont0=\sevenrm\rm}
\normalbaselines\rm}


\mathchardef\alpha="710B
\mathchardef\beta="710C
\mathchardef\gamma="710D
\mathchardef\delta="710E
\mathchardef\epsilon="710F
\mathchardef\zeta="7110
\mathchardef\eta="7111
\mathchardef\theta="7112
\mathchardef\kappa="7114
\mathchardef\lambda="7115
\mathchardef\mu="7116
\mathchardef\nu="7117
\mathchardef\xi="7118
\mathchardef\pi="7119
\mathchardef\rho="711A
\mathchardef\sigma="711B
\mathchardef\tau="711C
\mathchardef\phi="711E
\mathchardef\chi="711F
\mathchardef\psi="7120
\mathchardef\omega="7121
\mathchardef\varepsilon="7122
\mathchardef\vartheta="7123
\mathchardef\varrho="7125
\mathchardef\varphi="7127



\def\beginlinemode{\endmode
  \begingroup\parskip=0pt \obeylines\def\\{\par}\def\endmode{\par\endgroup}}
\def\beginparmode{\endmode
  \begingroup \def\endmode{\par\endgroup}}
\let\endmode=\par
{\obeylines\gdef\
{}}
\def\singlespace{\baselineskip=\normalbaselineskip}

\def\oneandahalfspace{\baselineskip=\normalbaselineskip
  \multiply\baselineskip by 3 \divide\baselineskip by 2}
\def\doublespace{\baselineskip=\normalbaselineskip \multiply\baselineskip by 2}

\nopagenumbers
\newcount\firstpageno
\firstpageno=2
\headline={\ifnum\pageno<\firstpageno{\hfil}\else{\hfil\elevenrm\folio}\fi}
\let\rawfootnote=\footnote		
\def\footnote#1#2{{\oneandahalfspace\parindent=0pt
\rawfootnote{#1}{#2}}}
\def\raggedcenter{\leftskip=4em plus 12em \rightskip=\leftskip
  \parindent=0pt \parfillskip=0pt \spaceskip=.3333em \xspaceskip=.5em
  \pretolerance=9999 \tolerance=9999
  \hyphenpenalty=9999 \exhyphenpenalty=9999 }
\def\dateline{\rightline{\ifcase\month\or
  January\or February\or March\or April\or May\or June\or
  July\or August\or September\or October\or November\or December\fi
  \space\number\year}}
\def\received{\vskip 3pt plus 0.2fill
 \centerline{\sl (Received\space\ifcase\month\or
  January\or February\or March\or April\or May\or June\or
  July\or August\or September\or October\or November\or December\fi
  \qquad, \number\year)}}


\hsize=6.5truein
\hoffset=0truein
\vsize=8.9truein
\voffset=0truein
\hfuzz=0.1pt
\vfuzz=0.1pt
\parskip=\medskipamount
\overfullrule=0pt	



\def\title			
  {\null\vskip 3pt plus 0.2fill
   \beginlinemode \doublespace \raggedcenter \bf}

\def\author			
  {\vskip 3pt plus 0.2fill \beginlinemode
   \singlespace \raggedcenter}

\def\affil			
  {\vskip 3pt plus 0.1fill \beginlinemode
   \oneandahalfspace \raggedcenter \sl}

\def\abstract			
  {\vskip 3pt plus 0.3fill \beginparmode
   \doublespace \narrower ABSTRACT: }

\def\summary			
  {\vskip 3pt plus 0.3fill \beginparmode
   \doublespace \narrower SUMMARY: }

\def\pacs#1
  {\vskip 3pt plus 0.2fill PACS: #1}

\def\endtitlepage		
  {\endpage			
   \body}

\def\body			
  {\beginparmode}		

\def\head#1{			
  \medskip\vskip 0.5truein	
  {\immediate\write16{#1}
   \raggedcenter \uppercase{#1}\par}
   \nobreak\vskip 0.25truein\nobreak}

\def\refto#1{$^{#1}$}		

\def\references			
  {\head{References}		
   \beginparmode
   \frenchspacing \parindent=0pt \leftskip=1truecm
   \parskip=8pt plus 3pt \everypar{\hangindent=\parindent}}

\gdef\refis#1{\indent\hbox to 0pt{\hss[#1]~}}	

\gdef\journal#1, #2, #3, 1#4#5#6{		
    {\sl #1~}{\bf #2}, #3 (1#4#5#6)}		

\def\refstylenp{		
  \gdef\refto##1{ [##1]}				
  \gdef\refis##1{\indent\hbox to 0pt{\hss##1)~}}	
  \gdef\journal##1, ##2, ##3, ##4 {			
     {\sl ##1~}{\bf ##2~}(##3) ##4 }}

\def\refstyleprnp{		
  \gdef\refto##1{ [##1]}				
  \gdef\refis##1{\indent\hbox to 0pt{\hss##1)~}}	
  \gdef\journal##1, ##2, ##3, 1##4##5##6{		
    {\sl ##1~}{\bf ##2~}(1##4##5##6) ##3}}

\def\endreferences{\body}

\def\figurecaptions		
  {\endpage
   \beginparmode
   \head{Figure Captions}
}

\def\endpage			
  {\vfill\eject}

\def\endpaper			
  {\endmode\vfill\supereject}


\def\ref#1{Ref.[#1]}			

\def\frac#1#2{{\textstyle{#1 \over #2}}}

\def\sla{\raise.15ex\hbox{$/$}\kern-.57em}
\def\leaderfill{\leaders\hbox to 1em{\hss.\hss}\hfill}
\def\twiddle{\lower.9ex\rlap{$\kern-.1em\scriptstyle\sim$}}
\def\bigtwiddle{\lower1.ex\rlap{$\sim$}}
\def\gtwid{\mathrel{\raise.3ex\hbox{$>$\kern-.75em\lower1ex\hbox{$\sim$}}}}
\def\ltwid{\mathrel{\raise.3ex\hbox{$<$\kern-.75em\lower1ex\hbox{$\sim$}}}}
\def\square{\kern1pt\vbox{\hrule height 1.2pt\hbox{\vrule width 1.2pt\hskip 3pt
   \vbox{\vskip 6pt}\hskip 3pt\vrule width 0.6pt}\hrule height 0.6pt}\kern1pt}


%
\newcount\eqnumber
\eqnumber=0
\def\Eqno#1{\global\advance\eqnumber by 1
    \expandafter\xdef\csname !#1\endcsname{\the\eqnumber}
    \eqno(\the\eqnumber)}
\def\Eqref#1{\csname !#1\endcsname}
\def\bR{I\kern-.4em R}
\def\bP{I\kern-.4em P}

\newcount\startpage

\def\qmax{$Q_{\rm max}\ $}

%
\twelvepoint
\title
{\bf Prediction of large events on a dynamical model of a fault}

\vskip 1.truein

\centerline{\it S.~L.~Pepke,$^{\diamond}$ J.~M.~Carlson,$^{\diamond *}$
and B.~E.~Shaw,$^{*\dag}$ }

\vskip .5truein
\noindent
{$^\diamond$Department of Physics, University of California, Santa Barbara, CA
93106}

\vskip .5truein
\noindent
{$^{\dag}$ Lamont-Doherty Earth Observatory, Columbia University, Palisades, NY
10964}

\vskip .5truein
\noindent
{$^*$Institute for Theoretical Physics, University of California, Santa
Barbara, CA 93106}

\vfill\eject

\vskip 1.truein

\abstract
\doublespace
\noindent
We present results for long term and intermediate term
prediction algorithms applied to
a  simple mechanical model of a fault.
We use long term prediction methods based, for example,
on the distribution of
repeat times between large events    to establish a benchmark
for predictability in the model. In comparison,
intermediate term prediction techniques,
analogous to the pattern
recognition algorithms CN and M8 introduced and studied by
Keilis--Borok et~al.,
are more effective at predicting coming large events.
We consider the implications of several different    quality functions $Q$
which can be used to optimize the algorithms with respect to
features such as space, time, and magnitude windows, and
find that our results are
not overly sensitive to
variations in these algorithm parameters.
We also
study the intrinsic uncertainities
which are associated  with
seismicity catalogs of restricted lengths.

\vskip .5 truein

%
%
%
\doublespace

\vfill\eject
\endtitlepage
\singlespace

\beginsection{I.  Introduction}

Prediction of the occurrence of
large earthquakes within a narrow space-time window
on a fault has proven to be a difficult problem for several
reasons:  i) time scales over which reliable and detailed seismological
records are available are often small compared to recurrence times within a
fault zone, ii) complexity of fault geometry and dynamics leads to great
variability in premonitory phenomena, iii) initiating mechanisms for large
events are not completely understood (which inhibits the determination of the
relative importance
of various precursors), and iv) knowledge of the strain distribution and yield
points along faults is insufficient to indicate the locations of future
epicenters.
The above make it difficult not only to predict well, but also to determine
how inherently predictable the system is and to find optimal forecasting
methods.

With relatively little certain knowledge concerning the system, one must be
concerned with optimizing prediction using the data at hand as well
as objectively evaluating the quality of the predictions that are made.
In seismology, these goals cannot be met using seismicity catalogs alone,
because  they  represent only a rather  brief record of the
system relative to the time scale of the seismic cycle.
One possible path forward is
through the use of artificial catalogs to compare methods of forecasting, since
one may generate a wealth of statistics for them on the computer and also have
the ability to vary system parameters.

In this paper we present results for long term and intermediate term
prediction  algorithms applied to catalogs generated
from a dynamical model of a fault.
Our objective is
not to prove that a particular
model will quantitatively duplicate the complex seismicity
patterns observed on real faults, though certainly
such an outcome would be of great interest.
Instead, this study will address
issues related to algorithm optimization
and the intrinsic limitations of algorithms given the sparsity of
data for the earth.
We consider
long term prediction techniques, such as the time-predictable
and slip-predictable
algorithms,
which are based solely on  characteristics
of the most recent large event.
In a similar spirit,
we also make predictions based on the distribution of time intervals
between large events, which can be determined to arbitrary accuracy on
a model such as that which will be considered here.
However, the principle results of this paper involve intermediate
term prediction techniques
analogous to those which have been developed by {\it Keilis--Borok et al.}
[1990{\it a, b}] and have recently been the object of
much attention (see, e.g., {\it Healey, et al.} [1992]).
The aim of these algorithms is to
provide an objective means for assessing the  probabilities
of large earthquakes
based on a collection of precursor functions, the values of  which
are determined by   regional
small and medium size events.
The precursor functions include
overall activity,
rate of change of activity, and clustering of events,
and they are
evaluated in coarse grained space-time windows.
Simple pattern recognition techniques
are then used to select the most relevant
precursors from a larger set of possibilities and to
establish threshold levels for signaling an alert.
On the model
we  consider the simplest versions of
these algorithms using only single precursors
and find that  they  do perform better than long term
prediction techniques.
However,
for the most standard precursors, such as the level of seismic activity,
the algorithms do not
work as well as we had hoped.
This still leaves open the possibility that
algorithms which utilize a combination of precursors  will be more
effective.
In addition, it is worth noting that one particular precursor function,
which is related to the degree to which
seismicity extends throughout the region,
does significantly outperform the other precursor functions we have considered,
and leads to  fairly reliable predictions on the time scale which
is relevant for intermediate term prediction.

Synthetic catalogs have been used a
great deal in the past couple of decades.  Usually, they provide a testing
ground for algorithms designed for things such as foreshock and aftershock
identification, as well as prediction.  As far as we know, to date
synthetic catalogs
have not been
used as a means for improving prediction algorithms, despite the
obvious benefits of synthetic catalogs, which include
better statistics and well-understood quantified properties.
These features allow for algorithm development and the study of
optimization procedures to an extent which is not possible for real catalogs.
  In
the face of little theoretical understanding of the problem of prediction in
such a complex system as the earth's crust, volumes of literature have been
published documenting possible precursory phenomena and event distributions.
Yet little is written on how  best to  use such information.  It is with
respect
to this question of optimal use of available information that synthetic
catalogs may prove most useful.

It is important to distinguish the two main types of
artificial catalogs, each of which provides a means to a different end.
First, there are purely  statistical catalogs,
which are constructed to satisfy  certain  statistical
constraints, such as consistency with
the Gutenberg--Richter law,  Omori's law, and/or spatio--temporal clustering.
However, because these catalogs are not  based on an underlying
physical process, as a test for predictability
they are most useful
to provide lower bounds on the effectiveness of
an algorithm.
A good algorithm should detect some inherent correlation
which has not been put in by hand and, thus, should do better  on the
earth than it does on any statistically generated catalog.

In contrast, one can consider artificial catalogs  generated by
dynamical models as we do here.
We will use
the model-generated catalogs as a means to the end of algorithm
optimization,  letting the ${\it physical}$
mechanisms guide us in determining which properties of the catalogs are
important for prediction. Here, unlike the purely statistical catalogs,
no features are $a$ $priori$ built in.  Our goal is to
identify features which are generic to a class of physical models and discover
how those features function in prediction algorithms which may be easily
adapted to different fault systems worldwide.

The model which we consider is a one-dimensional homogeneous
model for a fault,
which has recently been studied in a variety of  different
contexts including the statistical analysis of intrinsic scaling laws
[{\it Carlson, et al.}, 1991],
and applications to  dynamical fracture [{\it Langer and Tang}, 1991;
{\it Langer}, 1992].
The model
is a particularly good candidate for studies of seismic phenomena
because many
of its fundamental features
are reminiscent of behavior which is observed
in the earth. For example, the  magnitude vs.~frequency distribution is
similar to what is observed for a single fault or narrow fault zone [{\it
Carlson and Langer}, 1989{\it  a, b}],
and the model generates a moment
spectrum similar to those inferred from  seismographic
observations [{\it Shaw}, 1993{\it a}].
The model is also a particularly good candidate
for studies of predictability because,
firstly, it is deterministically chaotic, and hence technically unpredictable
at long enough times.
Secondly, the model exhibits
a sharp distinction between small and large events. The smaller,
more numerous events tend to cluster in the neighborhood of an
epicenter of a future large event [{\it Shaw et al.}, 1992]. This local
increase
in activity is a generic precursor in the model (other precursor functions
will also be considered) and is the primary
statistical basis for predictability at shorter time scales.
While the analogous behavior  is much less systematic in the
earth [{\it Kanamori}, 1981],
a similar rise in regional activity has been observed on
some occasions prior to large events, and
is one of the signals used for prediction in the algorithms such as CN and M8
which were introduced  and studied by
{\it Keilis--Borok et al.,} [1990 {\it a, b}].
The last important feature in the   model is its simplicity,
which allows us to numerically generate the equivalent of millions of years
data with perfect detection of events.
This feature allows us to evaluate quantitatively  the success
of algorithms in a manner that is impossible for real catalogs.

Our model
has the additional advantage of requiring no fine-tuning of parameters.
In particular,
there is a minimal number of  input parameters, all of which
are physically meaningful, yet with respect to which the qualitative catalog
features are robust over a wide range of values.  Further, we expect to be able
to predict events on the model better than on the real system, but that is to
the point:  to establish predictability limits for these nonlinear dynamical
models and to optimize on a ``clean'' system.  We hope that the things we learn
by doing so will shed light on the roles of various phenomena occurring during
the preparatory period for a large event and allow us to distinguish correctly
between meaningful causal premonitory characteristics and misleading
happenstance trends.  Ultimately, the prediction algorithms will be tested on
more sophisticated versions of the  model which may include aftershocks, a
coupled fault geometry, or be embedded in a two-dimensional medium.

The organization of the paper is as follows.  In Section II relevant
characteristics of the uniform Burridge--Knopoff (UBK)
model are reviewed.  This includes descriptions of
pertinent length and time scales which play a role in the model's
predictability.  In Section III we introduce the quality functions $Q$,
which are the means by which we evaluate the success of the algorithms.
In Section IV we present results for long term prediction, including
the slip-predictable model, the time-predictable model, and
prediction based upon recurrence intervals.  Such methods are
perhaps  the most widely
established procedures for determining earthquake probabilities along faults.
Section V contains our  results for
prediction based upon intermediate term precursors.  The forecasting method is
outlined and applied to individual activity-based precursor functions on the
model.  Section VI
discusses the robustness of our
results with respect to
variation of  the algorithm and catalog  parameters, including
variations in catalog length.
Section VII gives a summary and addresses outstanding problems.

\beginsection{II. Relevant Model Characteristics}

In the finite difference approximation,
the model considered here is one of a class first applied to earthquake
dynamics by {\it Burridge and Knopoff}  [1967].  It was reintroduced in its
simplest form and
analyzed in a modern context by {\it Carlson and Langer}
[1989{\it a, b}].
The one-dimensional
uniform Burridge--Knopoff (UBK) model represents the motion
of one side of a lateral fault
which is  driven by a slow shear deformation
relative to  the other side of the fault and
which is subject to a velocity--weakening slip-stick friction law at
the interface.
The system consists of ${N}$ blocks.
Each block is coupled to its nearest neighbors
with coil  springs representing the linear elastic response
of the system to compressional deformations.
A leaf spring attaches
each block  to a fixed upper surface, and  represents the
linear elastic response of the system to shear deformations.
The blocks
are constrained to move on the surface of  the sliding plate.
The system is loaded slowly by moving the lower plate at a velocity $\nu$
until one of the blocks exceeds the frictional  threshold.
That block then begins to slide, dissipating energy as determined by the
friction law.  The initially negative slope
of the friction law
leads to the essential dynamical instability which maintains the complexity
of the series of events over arbitrarily long time periods.  An event is
considered over when all of the blocks have come to rest.

In the continuum limit the partial differential
equation describing slipping motion
along the fault is:
$${\ddot U =  {{\partial^2 U} \over {\partial s^2}} - U - \phi [
\dot U+\nu]}\Eqno{pde}$$
where ${U(s,t)}$ is the displacement
measured with respect to the fixed
upper plate
as a function of position $s$ and time $t$.
Dots denote derivatives with respect to the dimensionless
time $t$, which has been scaled by the characteristic slip time for a
(homogeneous) large event.
Displacements $U$  have been scaled by the corresponding characteristic
slip distance.
In Eq.~(\Eqref{pde}) lengths $s$ are measured in units
of a stiffness length which is given by
the distance a sound wave travels (of order ten kilometers)
in the characteristic
slip time $\Delta t=1$.
In the finite difference version of (\Eqref{pde}), the
coefficient of the lattice Laplacian is
the number of blocks $\ell$ in this characteristic length,
so that the equilibrium block spacing
$1/\ell$ is the small scale cutoff.
This parameter plays a role in the dynamics of the UBK model (see, e.g.,~{\it
Langer and Tang}, [1991]),
and recent measurements  of microearthquakes suggest that there may be
a small scale cutoff in the earth as well. In particular, the measurements
of changes in the distribution of sizes of very small events
[{\it Malin, et al.}, 1989; {\it Aki}, 1987]
and measurements which suggest that the smallest earthquakes
may have a nearly constant rupture area (see, e.g.,
{\it Bakun, et al.} [1976], and
{\it Archuletta et al.} [1982]),
lead to a small length  cutoff  ranging from
meters to a few hundred meters.
This suggests that in units of the characteristic stiffness length
(roughly 10 kilometers as mentioned above) realistic values of
$\ell$  are  of order
$10^2-10^4$, i.e.~a large number.
Here we will take $\ell=10$ for numerical convenience. In
{\it Carlson et al.} [1991] the scaling of the UBK model as a function
of $\ell$ was considered.
The parameter $\nu$ in (\Eqref{pde}) is the dimensionless pulling speed,
given by the ratio of the rise time to the loading period
between large events. Realistic values of $\nu$ are thus $10^{-8}$
or less.  Here we will take $\nu$ to be small enough to
preserve the separation of time scales between individual events and
the loading mechanism.
For technical reasons when we do this we must also introduce the small
parameter $\sigma$ into the velocity-weakening friction law which we
will take to be
$${\phi(\rm z)= \cases{(- \infty ,1\rbrack,&\rm $z =0$; \cr
(1- \sigma)/ \lbrace 1+ \lbrack 2\alpha {\rm z}/(1- \sigma) \rbrack
\rbrace,&$z>0.$\cr}}\Eqno{frl}$$
Here $\sigma$ replaces $\nu$
in setting the scale for the displacement of the smallest events,
and we can set
$\nu=0$ while
the blocks slip.
The friction parameter
$\alpha$ is the ratio of the characteristic slipping speed to
the speed at which the friction is reduced by half the difference
between the threshold value and the value it attains at high
speeds.
It is difficult, if not impossible,
to  determine realistic values of $\alpha$ from
laboratory measurements.
For a wide range of $\alpha$ the behavior of the UBK model is not particularly
sensitive to the exact value.
Throughout most of this paper we will take $\alpha=3$ which is
in this range.
As pointed out by {\it Vasconcelos, et al.} [1992] and in
{\it Carlson, et al.} [1991],
for $\alpha$ small enough the behavior does change substantially,
though  we believe that this regime is less relevant to
seismicity.
The results which we present in this paper are relevant  in the
regime where $\ell$ is large, $\sigma $ is small, and $\alpha$
is large enough to generate the  generic  behavior.
As discussed above,
this regime is most relevant for seismological applications.

It is important to note that  the UBK model is homogeneous in
all of its material properties.
We observe complex behavior as
a consequence of a dynamical instability associated
with the friction law.
Beginning with a small heterogeneity in the initial condition,
we allow the system to evolve through several loading cycles
until it reaches a statistically steady state, at which point
the statistical properties are independent of the
details of the initial conditions.
In Fig.~1 we plot a small fraction (both in space and time)
of the catalog which will be used in this paper, which begins after the initial
transient period has passed.
For each event,
a line segment is drawn through all of the blocks that slip, and a cross
marks the position of the epicenter for each large event.
While there are clear correlations
at shorter time scales, because of the underlying homogeneity of the UBK
model the long time
average of the locations of epicenters of large events is independent
of position-- an event could happen anywhere with equal probability.

The behavior of this system is found to resemble that of an earthquake fault in
several important respects. Defining the  seismic moment $M$ to be the
total slip during an event:
$$M=\int_{\rm event}\delta U(s) ds\Eqno{moment}$$
and the magnitude $\mu=\ln M$, we find that
for wide ranges of the above
parameter values,  for small to medium size events the UBK model generates
frequency-magnitude distributions described by
the Gutenberg--Richter law
${D(\mu)=Ae^{-b \mu}}$. Here
${D(\mu)d\mu}$ is the frequency of
occurrence  of  events in the magnitude
interval $[\mu,\mu+d\mu]$
per
unit length per unit time (see Fig.~2).  As long as $\alpha$
is sufficiently large (${\alpha \ge 2.5}$), we obtain ${b=1}$ robustly in
this region.
In contrast, the largest events follow a different distribution which indicates
 an overfrequency of these events relative to the frequency extrapolated
from the distribution of small events.
The large  events are responsible for nearly all of the moment release
in the UBK model.
The overfrequency implies that there will be a characteristic repeat
time between large events, as discussed by {\it Carlson} [1991],
which can be
exploited for the purposes of long term prediction.
The change in behavior between the small and large events is characterized by
the length ${\tilde \xi \equiv {2  \over \alpha} \ln ( {4\ell^2  \over
\sigma } )}$.
In {\it Carlson and Langer} [1989{\it b}] it was shown that events which are
triggered in regions
of size less than $\tilde\xi$ tend to remain localized, that is,
the slip pulses which are generated
will decay rapidly when they encounter regions which are
far from threshold. In contrast, when the initial triggering zone
is larger than $\tilde\xi$,
the slip pulses will tend to propagate much further.
The length scale $\tilde \xi$ coincides well with the upper bound on
the clustering of small events such as those illustrated in Fig.~1, and the
associated crossover magnitude $\tilde\mu=\ln(2/\alpha)$
coincides with the minimum in the magnitude vs.~frequency
distribution (Fig.~2).
Roughly speaking, the
smaller
events  smooth the spatial configuration on length scales
less than ${\tilde \xi}$, thus preparing a region for the triggering of a
large
roughening event.
This pattern leads to  spatio-temporal clustering of small scale activity
along the fault  prior to a large event, as reported by {\it Shaw, et al.},
[1992]. While this systematic  increase in activity    is
a key feature leading to the relative
success of the intermediate term prediction
algorithms on the UBK model, this observation alone
is not sufficient to determine whether
ultimately the UBK model will be more or less predictable than the earth.

Throughout this work  we distinguish between
``small''
premonitory events and the ``large'' events we wish to forecast. This
distinction can be made precisely in the UBK model:
small events are
taken to be those for which ${\mu < \tilde \mu}$ while large events have ${\mu
\ge \tilde \mu}$.
While the existence of a
sharp feature is useful here to obtain quantitative results,
the relatively small number of events of size near
$\tilde\mu$ compared to events of lesser or greater magnitude,
implies that our results will not depend strongly on the exact criterion
for the crossover that is used.
This
division  is  certainly much less precise
in real data.
It is interesting to note that
when data from an individual fault or narrow fault zone is considered,
as in the UBK model
an  overfrequency of large events is observed [{\it Wesnousky,
et al.}, 1983; {\it Schwartz and Coppersmith}, 1984; {\it Davison and
Scholz}, 1985].
In these cases, typical magnitudes of the  large events, and geodetic
measurements of the plate rates can be used to estimate the recurrence
time interval, and make long term predictions.
In comparison, data accumulated over a broad region typically does
not show an overfrequency of large events. This is a consequence
of the fact that faults of many different sizes contribute to regional
seismicity. In this case,
recent results by {\it Pacheco, et al.} [1992] suggest a bend, or change in
$b-$value reflecting an underfrequency of large events,
in the magnitude vs.~frequency distribution of California
seismicity  at  roughly magnitude 6, which coincides
with estimates of the magnitude of events which just
span the full depth of the seismogenic zone.
These types of regional seismicity catalogs are typically used in
intermediate term prediction algorithms such as CN and M8 outlined in Section
V.

Simplifications which are inherent in
the  UBK model used here are its low dimensionality,
single fault dynamics, and lack of aftershocks.  Systems of interacting faults
and different fault geometries may ultimately
prove interesting but would be  most effectively represented in the context
of a fully   two-dimensional  elastic medium.
Studies of  a homogeneous two-dimensional  model are currently under way.
In addition, {\it Shaw} [1993{\it b}]
has proposed an aftershock mechanism
which if added
to the  two-dimensional model would lead   to
additional precursory phenomena.
Ultimately, we plan to study modified models in the context of prediction as
well.

\beginsection{III. Evaluation of Forecasting Algorithms}

In a spatially extended dynamical system, a prediction typically consists of
a projection, based on the current status of the system,
of the time and location of a coming event.
For example, when a hurricane is detected in the ocean,
meteorologists attempt to predict the precise time and location
at which the storm will contact the coast.
In comparison, the forecasts which are made
in seismology are necessarily more primitive and  involve much longer time
scales.
Rather than predicting how far in the future an event is likely to take place,
one forecasts the likelihood of an event occurring between
now and the end of some alarm period.
The time windows for such alarms can range from days for short term
predictions which are used
to alert the public and emergency rescue crews,
to   years for intermediate term predictions  which are used
to set certain guidelines for insurance and the
allocation of public resources,
and even tens of years for long term predictions which are   used to
establish  regional building codes.
In  algorithms CN and M8  one signals an alert
or {\it time of increased probability} (TIP)  to forecast a large event
in a specified region in space and time.
The most effective algorithms will not issue TIPs unnecessarily.
The TIP begins  when the algorithm detects that the system is in a
state of readiness, and, in order to minimize the cost associated
with signaling an alarm,
one must wait as long as possible  prior to the event
before turning on  the alarm.

In this context it is worth noting that there are a couple
of technical distinctions
between   algorithms CN and M8 and the
intermediate term prediction algorithms which we will consider here.
First, in the intermediate term algorithms it will be our goal to predict
the $epicenter$
of the large events, since in the UBK model
the precursory seismicity is strongly correlated with  this location
(since the long term algorithms make use of only information associated with
the last large event, in that case we will attempt to simply predict the time
of the next event, rather than focusing on the location of the epicenter).
In  algorithms such as CN and M8 the spatial regions which are considered
are much larger than the size of the large event
so targeting the exact location of the future epicenter is much less
relevant
(in comparison
the optimal spatial regions for the UBK model
are  comparable, and in fact
typically somewhat smaller than the size of the  event to be predicted).
The second distinction is more significant.
That  is, in CN and M8 the alarm duration is initially fixed
to be a specific time interval, typically five years.
After this period  has passed, the status of the region is
reevaluated, and a decision is made regarding whether or not to extend
the alert.
In contrast, after each event
we  reevaluate the status of the alarms, and the
average alarm duration is determined by the threshold for signaling
an alert.
In a model such as the UBK model, where the  precursor functions
tend to increase monotonically in the neighborhood
of the epicenter of a future large event,
alarms will typically (but not always) stay on   until
a large event has occurred.
This  monotonicity in the UBK model
is reminiscent of predictions based on time interval
distributions,
where it only becomes more likely to have a large event the longer it
has been since the last one.
The fact that such systematic behavior is not generally seen in the
earth does not preclude the use of the UBK model to provide
diagnostic tests of prediction techniques.
The idea here is to test the same sort of prediction algorithms
that are used in seismology on a model in which the basic
physical mechanisms leading to the complexity  are well
understood. Our results will help determine which algorithms
provide useful information under controlled
circumstances, and thus may ultimately be most successful
at predicting events in the earth.

With this in mind
we now  discuss methods for evaluating the long term and
intermediate term prediction algorithms which will be considered in
subsequent sections.
Forecasting  necessarily involves tradeoffs between
desired outcomes.
On one hand, one would like the TIP to be on for a minimal amount of time,
but on the other, one would like it to be on at the time a large
event takes place.
Another consideration which might come into play is the
expense associated with issuing false alarms.
In general, the cost of signaling an  alert must be balanced against
the cost of being unprepared when the event ultimately takes place.
While deciding how to make this tradeoff is to a large degree
based on public policy rather than mathematics or physics, the decision
is essential to algorithm optimization.

The success curve, which is defined to be
the fraction of events predicted vs.~the fraction of time an alert
is on, provides one meaningful way to
evaluate an algorithm.
Here each point on the success curve
corresponds to a different value of the threshold
for signaling an alert. Such curves are useful
to compare different methods of forecasting
and thus will be considered extensively
in the remainder of the paper (see, e.g., Figs.~6 and 9).
However, in order to $set$ the threshold
that will ultimately be used, one must
define a function on such a curve which will select
a particular value of the threshold over the others.
One function which is commonly considered in the
context of
algorithms  CN and M8 is called the {\it success ratio} which is
defined at each point on the success curve to be:
$$S={{\rm fraction\ of\ large\ events\ predicted}\over{\rm
fraction\ of\ time\ the\ TIP\ is\ on}}.\Eqno{S}$$
Thus $S$ specifies the gain relative to purely random prediction,
for which $S=1$.

Using $S$ as a measure of the quality of predictions has some inherent
complications, as pointed out by {\it Molchan} [1991] in the context
of some time interval distributions.
For example, when the
alarm time  decreases more rapidly than the number of events predicted,
one finds  that $S$ diverges
as both the alarm time and the fraction of events predicted   approach
zero.
We often observe this behavior in $S$ when intermediate term prediction
algorithms are applied to the UBK model.
Systematic algorithm optimization becomes an ill-defined problem in
this case.

To solve this problem,
we introduce a quality function $Q$  which will be used to
evaluate the long term and intermediate term prediction algorithms
discussed in the following sections.
It is worth emphasizing that such
functions rate the quality of  predictions
independent of the specific catalog or algorithm
to which they are applied, and thus along with the
success curve can provide  means by which different
algorithms can be successfully compared.

The specific choice of a $Q$ function is somewhat arbitrary.
In fact, we have considered several which are discussed
in more detail below.
Desired features of the $Q$ function are
(1)  it should be as simple as possible, with meaningful
parameters, so that policy decisions are easily mapped
onto parameter settings,
(2)  the definition should be flexible enough  that
alternative considerations can easily be incorporated,
(3) it should provide a clear measure of when an algorithm
is performing well with respect to some simple
base measure such as random prediction
or doing nothing,
(4) it should avoid the pathological behavior as alarm time goes
to zero which we observe in $S$,
and (5) the definition should be robust to small changes
in algorithm parameters so that determining a maximum value of
$Q$ is not an overly delicate procedure.

To satisfy the above criteria,
we define the quality function  $Q$ in its most general form to be
$$Q=\sum_{i=1}^N A_ip_i,\Eqno{Qgen}$$
where the $A_i$ are constant coefficients.
We will consider two versions of $Q$, which have similar behavior.
In the first $Q=Q_P$ and  the $p_i$ are the probabilities of
a set of outcomes with $0\le p_i\le 1$,  whereas in the second
$Q=Q_R$ the $p_i$ are interpreted as rates, measured here relative to
the overall rate of large events.
{}From a purely scientific point of view,
the probability based function $Q_P$
is the most standard sort of
measure, and  except near the extreme values of the ${p_i}$'s
its performance is similar to
an analogous product based function of the form $\Pi {p_i}^{\alpha_i}.$
In $Q_P$ we choose to use a sum rather than a product
to preserve the symmetry between using the success or failure rate
of a given outcome, i.e.~$p_i \rightarrow (1-p_i)$ and $A_i\rightarrow
-A_i$, is preserved in $Q_P$ (adding or subtracting a constant from $Q$ is
irrelevant).
In contrast,
from a public policy point of view,
the rate based function $Q_R$
is potentially the most useful.
Here $Q_R$ may be interpreted as a  cost--benefit function,
and the coefficients  can be set according to the relative
costs of the set of possible outcomes.
In this case the linear form of $Q_R$ is essential, since
benefits and costs are typically linear functions;
doubling the rate at which events are successfully
predicted and doubling the rate at which alarms are issued
roughly doubles the benefits and costs,
respectively.
Linear cost-benefit functions have also been used
in algorithms applied to real catalogs [{\it Molchan}, 1991].
While for the purposes of this paper, the distinction between
$Q_P$ and $Q_R$
is primarily a technical point
(we will often refer to $Q$ rather than specifically to  $Q_P$ or $Q_R$
when results and discussions are equally valid for either of these measures),
in applications to the earth and public policy decisions
there may be a significant benefit
to choosing one of these functions over another.

There are three outcomes which  we will incorporate
into the  $Q$ functions  we will be
using: the benefit of a successful alarm, the cost of an unsuccessful
alarm, and the cost of maintaining an alarm.
In both $Q_P$ and $Q_R$ (since the rates are defined relative to the
rate of large events)
we can define $p_1$ to be the
fraction of large events successfully predicted, and
$p_2$ to
be the fraction of the total observation time for which alarms are declared.
In $Q_P$  we define
$p_3$ to be the fraction of the total number of alarms that are issued
which turn out to be false, while in $Q_R$,
$p_3$ is defined to be the average number of false alarms
issued per large event (a nonlinear function of $p_1$ and $p_3$ relates
the two different definitions of $p_3$).
In each case,
the penalty for false alarms is necessary to avoid
the same sort of pathological behavior referred to above for the success ratio
$S$ in which the    solution to the optimization problem
is ill-defined and leads to  \lq\lq flickering"
alarms which are alternately turned off and on at very short intervals
in space and time.

%

Finally we  consider the coefficients $A_i$  for the $Q$ functions
represented by  Eq.~(\Eqref{Qgen}).
Since multiplying $Q$ by a constant does not change our ultimate
conclusions,   we
can normalize all the $A_i$ so that $A_1=1$.  Without loss of
generality, then, for $Q$ linear in the three probabilities or rates that
we are
concerned with, we can define $Q$ to be
$$Q=p_1-\vert A_2 \vert p_2- \vert A_3 \vert p_3\Eqno{Q}$$
where the coefficient of $p_1$ is positive, since it is a benefit,
and the coefficients of $p_2$ and $p_3$ are negative, since
they are costs.
Taking the optimal values for each of the outcomes
(i.e.~successful prediction of all of the events
$p_1=1$, with negligible costs $p_2=p_3=0$)
we obtain the upper bound $Q\le 1$.
Alternatively, if one simply does nothing  then $p_1=p_2=p_3=0$,
and we obtain $Q=0$.
We can thus use  the $Q$'s to say something interesting
about an algorithm: given the costs
which have been specified, are
there outcomes of the algorithm having $Q>0$; that is, are there
strategies that are {\sl {better than doing nothing}}?

The relative  coefficients $\vert A_2\vert $ and $\vert A_3\vert $
must be specified
by the user, and,
thus, will typically be different in different applications.
The parameter  $\vert A_3\vert $
sets a tolerance on the   number of false alarms that may occur.
The cost of  a false alarm will  depend upon the action
which the alarm prompts.
In some cases
one might be willing to tolerate quite a few false alarms, as in
for example,  the case of
short term prediction, where
the relative cost of issuing a false alarm
is much less than the losses  that might
be spared by successfully predicting a large  earthquake.
In comparison, for intermediate term prediction one might have
a  lesser tolerance for false alarms because of the expense associated
with maintaining a state of readiness over extended time periods.
In the sections that follow we will
consider some of the
implications of different choices
of this coefficient on, for example, the  optimization procedure.
In some cases for simplicity we will take $|A_3|=1$ (in particular,
for $Q=Q_P$), and while we expect that this value is somewhat larger
than the value which would be used in practice,
we will see in Section V that even this choice is not unreasonable,
as it leads to a false alarm rate in the intermediate term
algorithms which is comparable to those obtained
in algorithms CN and M8.

The coefficient  $|A_2|$  measures the cost of maintaining
an alarm.
Clearly, as $\vert A_2\vert \rightarrow
0$ the best strategy is to leave the alarm on all the time, in which case
$Q=1-\vert A_2\vert $.
Similarly,
as $\vert A_2\vert $ increases,  eventually the best strategy will be
to do nothing, in which case $Q=0$.
Here we will take $\vert A_2\vert =1$ so that only algorithms which
do not maintain a constant state of alert
will have a chance of doing better
than doing nothing at all.

Ultimately,
the different goals of short term, intermediate term, and
long term prediction will play a major role in selecting
the appropriate  function $Q$ for a particular use.
In the coming sections we will use both the success curve
and the $Q$'s to evaluate
prediction algorithms applied to the UBK model,
and compare the effectiveness of these two methods.
The success curve yields a more general comparison, showing
the full range of behavior as the fraction
of  time occupied by alarms   is varied.
Consequently, this measure will be most
relevant to  assessments of the relative performance
of algorithms which are intended to operate on
different time scales, such as the long term and intermediate term
algorithms which will be considered,
where the most appropriate
$Q$'s would be different in the two cases.
However, both the success curve and $Q$ will provide useful
information for comparisons between different intermediate term
techniques. In particular, $Q$ yields information
about the optimal time scales  associated with different
intermediate term measures,
and provides the most direct assessment
of how well an algorithm might be expected to perform in practice.
Finally, we cannot explore  questions  related to short term
predictability,
because the associated
time scales are not present in the UBK model.

\beginsection{IV. Long Term Prediction:  Results Based on Recurrence Intervals}

Long term prediction methods are used to estimate earthquake hazards on a
time scale of order tens of years.
The simplest such schemes make use of
only the magnitude or time of occurrence of the last large event,
and are referred to as the time-predictable and slip-predictable
models for hazard assessment [{\it Shimazaki and Nakata}, 1980].
Various applications of both time-predictable [{\it Scholz}, 1985;
{\it Bakun and McEvilly}, 1984] and slip-predictable
[{\it Kiremidjian and
Anagnos}, 1984] models to real faults have been made.
Historical records and geological information are used to construct
plots of accumulated slip as a function of time for a particular fault
or fault segment. The analogous plot  is constructed for  the UBK model
in Fig.~3, where the results correspond to the accumulated slip
for one representative patch along the fault over a time interval which
is long compared to available catalogs for real earthquakes, but
short compared to the catalogs which we consider later for the UBK model.
Essentially all of the slip is associated with
large events, which  is a  feature that is common to
both the UBK model and real faults.
In the slip-predictable scenario,
the magnitude of the coming large event
is correlated with the
time since the last large event.
Here the basic assumption is that were an event to  occur
today it would relieve all of the accumulated strain.
When this is valid the upper corners of the staircase
in Fig.~3 should fall on a line.
In contrast, in the time-predictable case, the
time interval preceding the coming large event
is correlated with
the magnitude of the last
large event. In this case,  the assumption is that
there is some roughly constant threshold which the local stresses must
achieve before a large event will be triggered, and
the system must reaccumulate a slip   deficit comparable to that which
was relieved in the last large event before the next event will be
triggered.
If this were  valid the lower corners of the staircase in Fig.~3
would fall roughly on a line.  The best linear least-squares fit to both
models are shown in the figure.  Although neither works particularly well,
the time-predictable model works somewhat better than the
slip-predictable model  on this relatively short UBK model catalog.

In order to confirm
this more generally,
in Fig.~4 we
test the (a) time-predictable and (b) slip-predictable
models on a much longer
artificial catalog. We also  test for correlations between
(c) the moments of subsequent large events and (d) subsequent time intervals.
A  strong correlation would be indicated by a heavy concentration
of points along a well defined curve. A strict adherence to either the
time-predictable or slip-predictable model would  lead to a
concentration of points along a straight line in (a) or (b)
respectively.
The errors associated with a linear least squares
fit to the data indicate that
the strongest
correlation is observed in 4a which implies that the time-predictable model
works best. In fact, none of the others shows any significant correlation
at all.  In view of the threshold dynamics which govern the UBK model, it is
not surprising that the time-predictable model better approximates the
behavior we observe.  What is more surprising is that even in
this case the correlation is very weak.
The fact that
the points are broadly scattered even in the best case indicates
the weakness of these techniques for prediction, even in a
simple model.  Significantly, however, for short enough catalogs one may
find much better correlation with one of the above long term prediction
models than actually exists when a sufficient amount of data has been
taken into account.  Regarding data from real faults, the case is indeterminate
due to the few number of data points available for any individual region,
as well as problems associated with accurate slip calculation during an
event.
(See, for example, {\it Thatcher}, [1984].)  The most reliable data seems to
favor time-predictability, however, and the USGS Working Group on California
Earthquake Prediction  (WGCEP) incorporates this
model into its long-term seismic hazard estimates [{\it WGCEP}, 1988, 1990].

Another method of long term prediction
which has been studied extensively in connection with
real
earthquakes is the use of probability distributions of  recurrence times for
large earthquakes on individual faults or fault segments.
These are used extensively by agencies such as the USGS [{\it WGCEP}, 1988,
1990]
to estimate the
probability of a large event over some  time period, say 30 years,
given the time of occurrence of the last large event.
The difficulty in this method lies in determining the correct distribution,
given the sparsity of data for large events on a given fault.
By combining data from many different faults, {\it Nishenko and Buland} [1987]
obtained a reasonably  good fit to a lognormal distribution.
Others [{\it McNally and Minster}, 1981] have argued that a Weibull
distribution is most appropriate.
While it is unlikely that the distribution will ever be known exactly,
a better understanding of the constraints  would be
useful because  the hazard assessments often rely on features
which are several standard deviations away from the
mean repeat time.
In fact, {\it Jackson and Davis} [1989] showed that, given
the sparsity of data, and the uncertainty in the recurrence time interval
distribution, large deviations  can
substantially alter the estimates for future earthquake
hazards, and in some cases lead to a projected
decrease in the earthquake hazard estimate  for  regions which exhibit
a  longer than expected gap since the last large event.

By comparison, the corresponding distribution can be determined
to essentially arbitrary accuracy for a model such as the UBK model. This was
done in the case of a  short fault (in which large events spanned
the entire system) by {\it Carlson} [1991]. The corresponding
distribution  is illustrated in Fig.~5a for the long fault catalog
that we consider here.
The best fit to  Gaussian (restricted to positive time intervals),
Weibull, and lognormal distributions are also
shown for comparison for the cumulative distribution in Fig.~5b.
Both  the Gaussian and Weibull
fit the distribution reasonably well, however,
the Gaussian does slightly better.
This is primarily  due to the fact that the Gaussian somewhat better
approximates the non-negligible weight at very short times
in  Fig.~5 which  arises from temporal correlations between large
events in neighboring  (and, in fact, slightly overlapping) regions.
By comparison the lognormal provides a substantially worse fit.
As for real earthquakes, in the case of the UBK model the standard
deviation is of order the mean repeat time,
with $\sigma/\overline T\approx .36$ for the model. This implies that even
if the distribution is known quite well, large uncertainties
will be inherent in long term prediction schemes.  This is discussed in
some detail by {\it Ward} [1992] in a similar application to a synthetic
catalog generated from a segmented fault model.  In
that case a Weibull fit to the cumulative probability distribution
proved superior to a lognormal (Gaussian was not tested),  with a
width which is roughly a factor of two
greater than that found in the UBK model.

In order to  quantitatively
compare our results for time interval prediction with the
corresponding results for the intermediate term prediction algorithms
discussed in the next sections,
it is useful to evaluate this scheme in terms of the  success curve
discussed in Section III.
In the most conventional method of time interval based prediction,
given some presumed distribution and known time since the last large event,
the probability of a large event in, say, the next thirty years is estimated
[${\it WGCEP}, 1988, 1990$]. In contrast, here we
specify a threshold time $t_0$ since the last large
event in a region
and  turn on an alert, or TIP, in that region
once a time $t_0$ has passed.
The TIP is turned off once the large event has occurred.
For each value of $t_0$ we then calculate the  fraction of earthquakes
predicted and the fraction of time the TIP was on.
In Fig.~6 the resultant success curve
is shown.  The data points correspond to  the cumulative results
for predictions which are made
independently
for each local position in space, and may be calculated directly from the
distribution of time intervals in Fig.~5a.
This represents an upper bound on prediction based on recurrence
intervals alone, because it incorporates the most detailed
spatial information.
In comparison,
for the purposes of long term prediction spatial
information is lost, resulting in a suppression in the success curve,
when the fault is coarse grained in the  manner
which leads to better and  more reliable intermediate term predictions.
In particular,
the  lower dashed curve corresponds to predictions which are made
in coarse grained regions of length
$3\tilde\xi$, as used
in the intermediate term prediction algorithms to follow.
The coarse graining of the fault
results in greater weight at very short times in the calculated recurrence
interval
distribution. This is due to
events which break through only a few blocks at
the edge
of a region, leading one to conclude the whole region is unstressed when,
in fact,
the probability of its sustaining a large event is still quite high.

The results can also be used to  evaluate the functions $Q$
in Eq.~(\Eqref{Q}).
Note that for time interval prediction the alarms are
only turned off after an event takes place so that the number of false alarms
is  zero ($p_3=0$). Hence
the probability based  function $Q_P$ and the cost-benefit function
$Q_R$ are identical, each  given by the difference
between the fraction predicted $p_1$ and the fraction of time the TIP is on
$p_2$ :
$Q=p_1-p_2$.
In Fig.~7 we plot $Q$ as a function of $p_2$.
Note that $Q$  takes a maximum value of \qmax=.46 (for the local
predictions) when the TIP is on
roughly $1/3$ of the time.
This corresponds to a  threshold time
which is $3/4$ of the mean recurrence time.
In comparison, \qmax=.36 for the coarse grained fault.

In the next section we will see that the
the success curve obtained here   falls below
the corresponding curves for most of the intermediate term
prediction algorithms, indicating that
if one is concerned solely with maximization of
the percent predicted while maintaining TIPs for a minimal amount of time,
then  the intermediate
term prediction methods perform substantially better than prediction based
upon recurrence intervals.
This is not surprising, given the breadth of the
time interval distribution (Fig.~5) and the clustering
of  small scale activity prior to large events which
we have observed in the UBK model.
However, time interval prediction is clearly  an improvement over
both random prediction and the option of
doing nothing at all. As a consequence, we expect that
this method  ultimately  will  play a role in
the optimal prediction algorithm for the UBK model.

\beginsection{V. Intermediate Term Prediction: The Pattern Recognition
Algorithms}

Intermediate term prediction algorithms are used to make earthquake
hazard assessments on the
time scale of one to five years.
Because forecasts are made
on relatively short time scales
compared to long term prediction,
more detailed information about the local state of the
system must somehow be deduced.
Regional small and medium size events provide one possible probe.
If a fault or fault segment is
near the threshold for slipping then one might expect that
the small scale seismicity
would also reflect the fact that the system is  close to
an instability. This sort of behavior occurs more generally
in a wide variety of complex systems. For example,
in  laboratory
fracture experiments,  the rate of microcrack production
accelerates prior to material  failure [{\it Mogi}, 1962]. In the UBK model,
we have
observed  an increase in the rate of  small to medium size events
prior to a large event. However,  while a local $increase$ in
seismicity has been observed prior to some large events,
and is in part the basis of some intermediate term
forecasts of, for example, the Loma
Prieta earthquake [{\it Keilis-Borok et al.}, 1990{\it c}],
this behavior is much less systematic in the earth.
In fact, in some
cases a local decrease in small scale seismicity,
or quiescence, is observed prior to a large event [{\it Wyss}, 1985],
and in others no change in the local
rate of seismicity is observed at all [{\it Kanamori}, 1981].
Herein lies the difficulty of intermediate term prediction.
The complexity of the earth yields many different activity patterns
so that it is difficult, perhaps impossible, to look at
one specific measure to make intermediate term forecasts worldwide.

For that reason {\it Keilis--Borok
et ~al.} [1990{\it a, b}] have developed pattern recognition algorithms,
such as CN and M8,
which can be applied to interpret objectively
the seismicity patterns in earthquake catalogs.
As  stated above,
the idea is that
small scale seismicity should signal a coming large event.
In these algorithms as
many as 18 different
possible precursory phenomena are considered,
and each precursor
casts a vote as to whether
or not an alarm should
be turned on.
The hope is that if a large event is not
preceded by a signal in a particular precursor, then it  might
be preceded by a signal in another.
Algorithms CN and M8  use seven precursor functions
and yield a success ratio (Eq.~(\Eqref{S})) of  roughly $S=4$.
In particular, roughly 80\% of events are successfully
predicted, when the TIPs are on 20\% of the time.
In comparison,  the value of $S$ for an individual precursor
function is typically of order $S=$2 or 3, corresponding to successful
predictions
of only 40\% to 60\% when the alarm time is 20\%.
While these values exceed what one would obtain
for purely random prediction $S=1$,
alarm times which are a smaller fraction of the seismic
cycle would be more useful for
intermediate term predictions.

On one hand,  algorithms such as CN and M8 are the best candidates for
objective
means to signal alerts, while on another, they remain somewhat speculative
because  a systematic evaluation using seismicity data
is impaired due to the limited amount of available data.
In fact, {\it Dieterich} (in {\it Healy et al.} [1992])  has suggested that
instead of comparing the success of the algorithm to
purely random prediction, one should  compare to the success
obtained when the TIPs are randomly set but with a bias  determined by the
local rate of large events, thus  building in an element of long term
prediction.
As we will show, even in the case of single
precursor functions the intermediate term algorithms
do  perform well in comparison to long term prediction
methods previously  discussed for the UBK model.
However,  for the most standard  precursor functions
the performance is still not as good as one might expect, and leads to
alarm times which are somewhat longer than desired for intermediate term
prediction.
There is one exception to this, which we will describe in more detail
below.
Ultimately we hope to provide an answer  to the question of whether
in general
a systematic improvement of these algorithms is possible for the earth.

Below we outline the pattern recognition algorithms which we will use.
While we will not consider algorithms which utilize multiple precursors in
this paper,
our algorithms are designed to closely mimic the simplest versions of the
pattern recognition algorithms such as CN and M8
which
have been studied extensively in the context of real
earthquakes by {\it  Keilis--Borok et al.} [1990{\it a, b}], and in
earlier model studies by  {\it Gabrielov et al.} [1990].
The algorithms require  precursor functions $\{ f_i(\Delta s,
\Delta t)\}$, where $f_i (\Delta s,\Delta t)$ is the $i$th
precursor function, evaluated in the spatial region
$\Delta s$ (these
are large overlapping
circles for the earth, and overlapping line segments
in the  one-dimensional UBK model which we will take to be a distance one
apart)
during a sliding time window $\Delta t$.
Throughout this section we will
take $\Delta s=3\tilde\xi$ and $\Delta t/\overline T=.36$ (i.e.~time windows
which are $36\%$ of the mean recurrence interval). In Section VI
we will show that the performance of the pattern recognition algorithms
employed here are not overly
sensitive to these choices.

The statistics which are used to evaluate the success curve
and quality functions $Q$ (see Section III) are compiled
individually within each of these spatial regions, and then
combined to determine the cumulative result.
For each region $R$, the events which are considered to be in $R$
are those events for which the $epicenter$ lies in $R$.
For the UBK model our goal is to predict the epicenter of the
large events.  The precursory seismicity is strongly  correlated
with the epicenter of  coming large events, and it is
this correlation which we expect the pattern recognition algorithms to
detect.
Thus the  fraction of large events which is successfully predicted
($p_1$ in Eq.~(\Eqref{Q})) in $R$ is the fraction of events
for which the alarm was on in $R$ when a large event with epicenter
in $R$ takes place.
Similarly,
the fraction of time an alarm is on ($p_2$  in Eq.~(\Eqref{Q}))
is also determined for each region individually, as is the fraction
(for $Q_P$) or rate (for $Q_R$)
of false alarms ($p_3$).
As previously stated,
unlike  algorithms CN and M8, here
we will not specify a fixed alarm time. Instead the
status of alarms will be reevaluated after each event.
Because our precursor functions tend to increase monotonically
prior to a large event,
once an alarm is turned on it will tend to stay on until an event takes place.
The average alarm time is
determined by the threshold value of the precursor function,
and a false alarm will  be counted if the alarm is turned off
before the region in question has contained the epicenter of a large event.
Thus  false alarms typically arise in two
fashions.
For relatively low values of the precursor function thresholds,
occasionally
an alarm may be turned off then reinstated prior to a large event.
Alternatively,
a region may experience the typical precursory upswelling of activity,
but then be preempted as the epicenter by an event which is triggered
in a nearby region (recall that the goal is to predict the epicenter).
The pathological flickering
behavior, in which alarms are rapidly turned off and on,
which for injudicious choices of $Q$ functions can lead to an
apparent maximum in $Q$ as the alarm time goes to zero, is
avoided by including
this penalty for false alarms with a reasonable choice of the coefficient
$|A_3|$.

For   algorithms CN and M8, precursor
functions   take a discrete set of values simply referred
to as high, medium, or low. A high value
(the threshold level \lq\lq high" is determined from existing catalogs)
casts a vote in the favor
of issuing a TIP.
In our case, we allow for a continuous distribution
of values, and
the most effective
precursor functions will minimize the overlap between the distribution of
values  taken near the time of a large event and the set of values
taken over all time.
Examples of both the distribution of values just prior
to a large event and the background distribution averaged over all
time
is illustrated in Fig.~8 for the activity precursor function $f_1$.
Here activity is defined to be the total number of small to medium
size earthquakes,
independent of magnitude, within a space-time window (with the exception of
events involving only a single block, which are omitted for convenience).

In Fig.~8 we see that the activity $f_1$ is typically small.
The large spike at $P(f_1)=0$ reflects the extended quiescent period which is
observed just after  a large event in the UBK model.
In contrast, the conditional probability of the value of
$f_1$ at the time of a large event $P(f_1|{\rm large\ event})$
has very little weight at $f_1=0$, because the  neighborhood of the epicenter
nearly always exhibits some activity just prior to a large event.
In fact, within the specified space-time window and for the parameter
values we have taken, the activity
before  a large event can reach values as high as a few hundred
events. From Fig.~8
we see that the average activity just before a large event is
roughly $f_1=30$. At that point
we observe  a ratio of  approximately five between
the conditional probability and the background
value, suggesting that activity should be a  good
precursor function.

In addition to (1) activity, we will consider the following precursor
functions:
(2) rate of change of activity $f_2$ defined to be the slope of a
linear least
squares fit to the activity as a function of time within a space-time
window,
(3) fluctuations in activity $f_3$ defined to be the
root mean square deviation from the linear least squares fit
used in (2), and
(4) active zone size $f_4$
defined to be the number of blocks that have slipped within the space-time
window, independent of how many times they have slipped.  Moment-weighted
activity was also examined, but found to differ little from activity.
While activity, rate of change of activity, and fluctuations
in activity all have direct analogies in seismicity catalogs,
the active zone size is more difficult to determine for real earthquakes.
We choose to consider it here  to broaden our set of possible precursors,
and, as we will show below, for the UBK model this precursor function
performs particularly well.  Note that this is the only precursor function
which is set to zero after a large event occurs (zeroing the activity in a
similar fashion affects the results very little).

Active zone size
is a measure of the extent to which  seismicity in a given
region is diffuse. It
is not simply a size or moment weighted activity
measure, nor is it a measure of clustering of events.
Instead, it is more directly a measure of the
broadening or anticlustering
of small to moderate size  events, which leads to the development of
a   nucleation region
associated with a coming large event.
To measure active zone size in the earth
a box counting algorithm might be used, in which the large spatial regions
taken in algorithms CN and M8 would be subdivided into many smaller
regions, and
the number of these smaller regions containing seismicity at or above a
certain level  would  define the regional  active zone size.
Of course, in the earth this measure is somewhat more complex due to the
variable complexity of fault networks in different regions in the
earth.
However, properly normalized to account for such differences,
it may be worth considering because of the exceptional
performance of this measure for the UBK model.
Compared to activity (Fig.~8) the signal to background ratio is
somewhat greater for active  zone size,
and is somewhat less for rate of change of activity and fluctuations in
activity.

Next we will implement the intermediate term prediction
algorithm individually for each of the above precursor
functions. As previously mentioned, this involves
coarse graining the fault into overlapping line segments of size
$\Delta s=3\tilde\xi$ a distance $\ell$ apart, and evaluating the
precursor functions $\{f_i(\Delta s, \Delta t)\}$ individually on the subset
of small to medium size events $(\mu\le\tilde\mu)$ which occurs
within each space--time window.  For each precursor
we vary
the threshold level for signaling a TIP, and monitor
the
resulting
fraction of the
large events  $(\mu>\tilde\mu)$ successfully predicted as well
as the fraction of time TIPs were on.
This  yields the
success curves
illustrated in Fig.~9.
Note that for each of the precursor functions  the algorithm leads to
an enhancement over random prediction and a clear
improvement over the results obtained in the last section
using long term techniques. The activity measure $f_1$,
which is most easily interpreted seismologically,   yields results for
the UBK model which are somewhat better than those quoted above for algorithms
CN and
M8, and is a significant improvement over the results obtained using
single precursors on real catalogs.
In comparison to $f_1$,
rate of change of activity $f_2$ performs somewhat worse, and
fluctuations in activity  $f_3$ performs significantly worse.
This is not too surprising in
light of the fact that in the UBK model the increase in activity is
essentially monotonic.
What is more surprising is the extent to which active zone size $f_4$
outperforms the others.
In that case nearly all of the events are successfully predicted when
the alarms are on only 5\% of the time.
Of those we have considered,
active zone size is the only measure which clearly leads to predictions
which are relevant on the time scales associated with
intermediate term prediction (of order 5\% of the recurrence
interval).

The corresponding $Q$ curves for both $Q_P$ and $Q_R$
are illustrated in Fig.~10, where it is clear that the results
obtained are comparable for the two measures.
According to these criteria, the active zone size provides the
highest quality (maximum \qmax) prediction in both cases.
In the UBK model the effectiveness of $f_4$ as a precursor
can be traced to the fact that
most of the energy dissipation occurs
during large events. In comparison, small to moderate
size events relieve relatively little stress. Thus when a small
event has occurred it is a direct signal that the blocks involved
are poised at the threshold of instability.
For a given region, therefore, the active zone size is a more
direct measure of the density of blocks that are close to threshold
than activity,
although these two quantities are clearly related.
The second highest quality predictor (for this choice of space-time windows)
is the activity, which, in fact, outperforms   active zone  size for
larger alarm time fractions.
Because it is most closely related to  quantities which are easily
calculated from seismicity catalogs,
in the next section   we will use activity as a base
measure with respect to which we will  optimize the algorithm.
In Fig.~10a for activity, based on the measure $Q_P$  with $|A_3|=1$,
we find that \qmax  occurs when the threshold value
$F_1=12$ is used. In this case, $p_1=.92$ is the fraction
of large events successfully predicted, $p_2=.14$ is the fraction
of time the alarm is on, $p_3=.48$ is fraction of false alarms
(i.e.~roughly 1/2 of all alarms that are issued do not result in
an
epicenter of a large event occurring in the region
during the  associated alarm time).
It is interesting to note that the threshold value $F_1=12$
does not coincide with the maximum in the conditional
probability illustrated in Fig.~8
indicating room for improvement when ultimately multiple predictors are
considered.
Finally, using this criterion
for evaluating predictions,
rate of change of activity  performs somewhat worse, achieving a broad, but
low maximum when alarms occupy roughly fifteen percent of the total time,
but still provides
a significant improvement over doing nothing.
In contrast, fluctuations  in activity
perform  significantly worse.

It remains a topic of current research to determine the extent to
which our results might be improved by combining the
different precursors.
For example, appropriate combinations of the precursor functions  may lead to
threshold functions which result in an increase of the number of
successful predictions
or a reduction in the false alarm rate and the amount of time
occupied by alarms.
This is illustrated in Fig.~11, where
each point corresponds to the simultaneous measurement
of
the activity  $f_1$ and the rate of change of activity
$f_2$  (a) over all time and (b) just before a large event.
Currently, for activity alone $Q_P$ is optimized when alarms are declared for
activity at or above the threshold level $F_1=12$.  Similarly,
for rate of change of activity alone at $Q_P=Q_{max}$ alarms are declared at
or above the threshold level
$F_2=120$.  At these values of $F_1$ and $F_2$ the set of events which are
predicted using $f_2$ is a subset of those which are predicted using $f_1$.
However, by comparing the background and conditional distributions, it is
clear that in some cases while $f_1$ maintains a relatively high value, $f_2$
has dropped far below the value it takes at the time of a large event.  Thus
by choosing a threshold function such that alarms are declared, for example,
only when
the values of $\it both$ activity and rate of change of activity exceed
some specified individual thresholds
one might find that
the number of false alarms and the total alarm time
could be significantly
reduced while the number of successfully predicted events
might decrease very little.
Such combinations  may improve the overall performance in terms of $Q$.
We are currently developing multidimensional  optimization techniques
with which we may address this problem in more detail and we hope
the results will be useful for further evaluation of the  pattern recognition
algorithms CN and M8.

\beginsection{VI. Intermediate Term Prediction:
Variation of Parameters}

The problem of pattern recognition leads naturally to a question of
optimization:  under what circumstances is the algorithm most likely to
correctly recognize a sequence of events indicating an imminent earthquake?
In the most general case, one might consider optimization
within the space of all possible algorithms, which make use
of an arbitrarily large
number of physical attributes (e.g.~precursors), and then also
optimizing each algorithm with respect to the input parameters which are
used.
Clearly,
this is an extremely high dimensional and difficult problem, which, if
feasible,
would be of great interest.

In the pattern recognition   algorithms such as CN and M8,  the basic features
of the
voting algorithm are fixed, and the problem of pattern recognition
refers to the selection of a subset of
precursor functions out of a specified list of possibilities
as well as optimization with respect to algorithm parameters, such as
threshold levels for individual precursor functions.
Thus, to make the optimization problem more tractable,
the dimensionality of the space
is reduced by first  deciding on a voting algorithm, and then using
specific knowledge of
the earthquake process
to select possible precursor functions, such as those considered in
Section V, which are  thought to be most relevant.
In this reduced space, the algorithms may be optimized with respect to
the remaining parameters.

In a similar manner, in this section we consider the question of
optimizing the algorithm studied in Section V  for the UBK model.
Unless explicitly stated otherwise,
we will restrict our attention to the probability based
function $Q=Q_P$, though some issues related to the
sensitivity of certain aspects of the optimization procedure to
the choice of $Q$ functions will also be addressed.
The question of restricted magnitude windows will also be considered.
In particular, we  maximize $Q(F_i,\Delta s,\Delta t,
\Delta \mu)$, where $F_i$ is the alarm threshold for the $i$th precursor
function $f_i(\Delta s, \Delta t)$, $\Delta s$ and $\Delta t$ are the
space and time windows within which the function $f_i$ is evaluated, and
$\Delta \mu$ is the magnitude window from which events are taken.  Finally,
we
consider stability of the maximization procedure as a function of catalog
length.

Although the four parameter optimization problem for $Q(F_i,\Delta s,\Delta
 t,\Delta \mu)$ is certainly doable in a brute force fashion (the computing
resources needed for such a calculation are not too large), it seems
sensible to maximize $Q$ with respect to $F_i$, $\Delta s$, and $\Delta t$
simultaneously in order to first optimize the space and time windows.
Then the
variation of \qmax with $\Delta \mu$ (once again allowing $F_i$ to vary) may
be studied separately.  Examining such cross-sections
allows for the  future possibility
of including the magnitude dependence
explicitly within the precursor function definitions, as is done in
CN and M8.

Below we will restrict our
attention to the activity
precursor $f_1$, because it is the most easily interpreted seismologically.

\noindent
{\it VI.a Optimization with respect to space and time window size}

The pattern recognition algorithms CN and M8 are typically evaluated with
time windows of order five years
and spatial windows of order
one thousand kilometers (the specific size scales with the moment of the
large event).
Once a TIP is alerted within a particular space-time window,
the system is sometimes reevaluated within the original  window
but  on a more finely
coarse grained catalog in space (roughly a few hundred kilometers)
and time (roughly one year).
While these secondary predictions are somewhat less reliable,
and do not work   well when applied  independently,
on occasion they  do serve to more closely pinpoint the target region
for a coming large event.

One interesting, and somewhat
surprising feature  of CN and M8 is the  effectiveness
of the large spatial
window sizes that are used for the primary predictions.
In particular,   spatial boxes  are typically set to
be an
order of magnitude
larger than the target event.
The notion that correlations might extend
over anomalously broad regions is akin to the
ideas of self-organized criticality introduced by {\it Bak, Tang, and
Wiesenfeld} [1987],
which postulate that  a  large
class of driven dissipative systems
may be attracted to
dynamical states  which display large correlations reminiscent of
equilibrium critical points, and that this behavior arises
due to  instabilities associated with threshold dynamics.  Such long-range
correlations are typically indicated by power law frequency spectra, such as
that embodied in the Gutenberg-Richter law describing fault systems.
However, because
statistics associated with individual faults or narrow fault zones do not
exhibit power law behavior over the whole frequency--magnitude
spectrum (instead one observes an over-frequency of large events
relative to small and medium-sized ones), we do not anticipate that individual
faults should be described within the context of a simple critical phenomena
theory.  It is also for this reason that we would not expect the UBK model to
exhibit correlations on scales as large as those used in CN and M8.

In Fig.~12 we evaluate $Q_{\rm max}(\Delta s, \Delta t,\Delta  \mu)$
for the activity precursor function $f_1(\Delta s,
\Delta t)$ as a
function of the spatial $\Delta s$ and temporal $\Delta t$
window size.
For each data point,
the optimization with respect to threshold level $F_1$ has already been
performed, and we include all of the small and  medium size events
$\mu_{\rm min}\le \mu\le\tilde\mu$, so that
$\Delta \mu=\tilde\mu-\mu_{\rm min}$ with
$\mu_{\rm min}$  fixed
at the minimum magnitude of a two-block event (the smallest event
we have retained in the UBK model catalog),  in our evaluation of the activity.
The fact that the surface contains a relatively
broad maximum, with no sharp features, is  an indication that
the algorithm is reasonably robust with respect to variations in the
parameters.
The fact that the maximum is roughly L-shaped  indicates that as
long as either the spatial windows or the time windows are taken to be
near the optimal size, there is reduced sensitivity to
variations in the other parameter.
We find that the optimal setting is
$\Delta s \approx 3\tilde\xi$ and $\Delta  t/\overline T \approx .36$.
For these  space-time windows, the optimal activity threshold
value is
$F_1=12$,
which leads to a value of  $p_1=.92$ for the fraction of events successfully
predicted, with alarms on $p_2=.14$ of the time, and a $p_3=.48$
rate of false alarms (i.e.~roughly one false alarm for every successful alarm).
The value of \qmax is thus \qmax=.30.

The natural spatial scale for correlations in the UBK model is $\tilde\xi$,
so it is not surprising
that the optimal box size  is of order this length.
Indeed taking boxes of size $\tilde
\xi$ and counting
the number of events per box (a box centered at each block) in
the time interval preceding a large event, it is found that the box with
maximal precursory activity lies within ${\tilde \xi}/2$ of the future
epicenter 80\% of the time.
Fig.~13a illustrates a slice
of an optimization surface (Fig.~12) along the spatial direction.
It is clear that $Q$
rises dramatically up to a length close to $\tilde \xi$ then becomes relatively
less sensitive beyond that length.
As discussed in more detail below, the coefficient $|A_3|$ for the false
alarm penalty in $Q$ plays an important role in setting the tolerance
for the
minimum window size for which reasonable values of  $Q$ are obtained.

Another length scale which is relevant to the problem of spatial optimization
is the typical size of a large event $\xi^*$. While we do not yet have
an analytical expression for $\xi^*$, in {\it Carlson et al.}
[1991]  numerical simulations were used to deduce that
$\xi^*\sim\ell\tilde\xi$.
Thus for the choice
of parameters considered here  we estimate that $\xi^*\approx 10\tilde\xi$.
While naively one might then expect the
algorithm to optimize at $\xi^*$
we
observe a broad peak in \qmax (see Fig.~13a)
at spatial
windows between  $3\tilde \xi$ and  $4 \tilde \xi$.
The fact that this  is somewhat less than $\xi^*$
reflects the increased alarm time for larger window sizes,
which, because activity
is only correlated with an  epicenter on length scales of order $\tilde \xi$,
does not result in an
increased
prediction of epicenters.
Thus
it is somewhat better  to use spatial windows which are
smaller than $\xi^*$
on the UBK model fault.
By comparison
this spatial window size is significantly smaller than  that used
in CN and M8.

The optimal time window of $\Delta t/\overline T=.36$ is relatively large,
corresponding to
time windows which are   $36\%$ of the mean recurrence interval.
One can see this in Fig.~13b which illustrates a typical slice of the
optimization surface (Fig.~12) taken along
the temporal direction.
As observed previously in {\it Shaw, et al.} [1992], there is a  significant
increase in  the rate of change of the cumulative activity
after on average 2/3 of the cycle time has passed.
Because the activity increase is essentially monotonic,
and  because we have made the alarm time  independent of the time window size,
there is no cost for having large time windows.
As a result the optimization over time  windows is essentially
picking out the activity \lq\lq turn-on" time.
Note that the optimal time window is relatively much longer than those
employed in algorithms CN and M8 (typically 6 years).

The optimal spatial and temporal windows for a few other values of the UBK
model
parameters were also evaluated (specifically $l=10, \alpha=2$ and $l=8,16,
\alpha=3$)
and found to be consistent with the above
results when
expressed in terms of the length $\tilde \xi$ (calculated using the relevant
$l$ and
$\alpha$ values) and the mean repeat time.

Finally, it is worth emphasizing that the large spatial and temporal windows
which are selected by our optimization procedure were not obtained
when we first attempted to optimize the algorithm  with respect to
the success ratio (Eq.~(\Eqref{S})) or a $Q$ function
which  neglects the penalty for false alarms ($A_3=0$ in Eq.~(\Eqref{Q})).
Instead, for example, in the case of the success ratio $S$
as the space-time windows were decreased
$S$ was found to increase essentially without bound.  The analog of this
behavior for the $Q$ which neglects false alarms is illustrated
in Figs.~13a and 13b.
As previously mentioned, this occurs because as the activity threshold
is increased, the alarm time  goes to zero faster than the fraction
of events successfully predicted, leading to
an ill-defined optimization problem.

The incorporation of a penalty for false alarms leads to
large space-time windows, and  greater reliability of the results.
However, it is worth noting that the optimal window size
is somewhat sensitive to the choice of the coefficient $|A_3|$
in $Q$ (Eq.~(\Eqref{Q})). For example, in $Q_P$ we find
that the onset of the rapid decrease in \qmax as the spatial  window
size is reduced (Fig.~13a)
occurs for windows which scale linearly with  $|A_3|$:
$\Delta s\sim 2|A_3|\tilde\xi$.
Thus while for large enough space-time window sizes,
the value of \qmax is  generally not very sensitive to the
size of the windows, the optimization procedure may lead to
smaller windows as the penalty for false alarms is decreased.
In fact, in Fig.~13a with the choice $|A_3|=1/3$, the optimal window size
is less than $\tilde\xi$, which  results in
preemptive false alarms being issued within the nucleating region
(of roughly size $\tilde\xi$)
of the coming large  event, simply because the windows were not taken
to be large enough (i.e.~the active window did not receive credit for
predicting the event since it  was too small to contain the
epicenter).
For this reason $|A_3|$ of order unity
is a natural choice for the probability
based $Q_P$.
Furthermore, while
qualitatively similar behavior  (lack of strong sensitivity of
\qmax over some reasonable range of  window sizes, with a
cutoff which scales linearly with the false alarm penalty $|A_3|$)
is observed for the rate
based cost--benefit function $Q_R$, the natural choice
for $|A_3|$ in that case will be less  because in that case
$p_3$ is a rate (and can thus exceed unity) rather than a probability.
For $Q_R$ we observe that the onset of rapid decay
in \qmax as a function of decreasing $\Delta s$
scales roughly
as $\Delta s\sim 10|A_3|\tilde\xi$.

\noindent
{\it VI.b Magnitude windows}

Next we consider the effect of restricting the range of magnitudes
which is used to evaluate the activity precursor function.
Restricted magnitude windows are an important consideration
in seismology because  typically reliable data is only available
over a relatively narrow range of magnitudes. Furthermore, in  CN and
M8 seismicity from different magnitude
ranges
is counted separately as different precursor functions.
Because they are more frequent,
small events have much more reliable statistics. However, they
also tend to swamp the medium
sized events in  measures such as total  activity.
By considering  only events larger than
specified cutoff magnitude  we  determine
whether there is a
statistically  significant increase in the   moderate size events, apart from
that
predicted by an extrapolation of the increased rate of smaller events which
occurs as
precursory phenomena.
An increased rate of medium size events prior to a large earthquake
has been observed in certain instances
in the earth (see, e.g., {\it Pacheo et al.}, [1992]).

In Fig.~14 we illustrate our results for \qmax as a function
of the lower magnitude cutoff $\mu_{\rm min}$.
In each case, we use the spatial and temporal windows sizes
$\Delta s=3\tilde\xi$ and $\Delta t/\overline T=.18$.
In addition, the upper cutoff $\mu_{\rm max}=\tilde\mu$ is fixed.
Variations in
$\mu_{\rm max}$ should not significantly alter the results, because the
total activity is dominated by the events at the lower cutoff.
{}From Fig.~14 it is clear that for a wide range of smaller values
of $\mu_{min}$, \qmax
is  quite insensitive to the lower magnitude cutoff. In fact, in
previous sections  we have used
this feature to ignore the numerous one-block events
in our catalog. These events are the most numerous, but  their retention
does not improve our results.
In contrast, for large values of
$\mu_{\rm min}$ there is eventually
a sharp (linear) decline in \qmax, which
is dominated by a decrease in the number of large events
predicted ($p_1$ in Eq.~(\Eqref{Q})).
It is clear that eventually a decline must be observed
once the rate of small to moderate events
that will be counted  becomes comparable to the overall  rate of
large events (see Fig.~(2)). This  corresponds to $\mu=-7.2$ for the parameters
we have chosen.
The onset of this behavior occurs for  a slightly smaller,
but comparable, magnitude
$\mu_{\rm min}=-7.5$
(the discrepancy is associated with the fact that the
algorithm is not generally optimized by setting the threshold
at the peak of   activity distributions such as Fig.~8).
This
coincides
with a threshold of unity for signaling a TIP.
There appears to be a slight maximum in \qmax just prior to
the decline. This suggests that in the UBK model
there may be some
additional precursory feature associated with medium size events,
and that  it may be useful
to consider the moderate size events as a precursor separate from the
small events as is done in algorithms CN and M8.  This feature is
more pronounced for time windows which are  smaller than the
optimal windows for the activity precursor
(the figure shows our results for $\Delta t/overline T=.18$
rather than the window $\Delta t/\overline T=.36$
for optimization with respect to
activity for this reason)
which reflects the fact that the largest precursory event also typically
occurs relatively close in time to the main event.

\noindent
{\it VI.c Variation of catalog length}

Finally we consider the effects of varying the length of the catalog.
Because real seismicity catalogs are short  compared to the
time scale of the seismic cycle (30 years of reliable data on small to
moderate size events while the cycle time is
of order hundreds of years), restrictions
on catalog length may significantly constrain the ability of any
seismicity based algorithms to predict.  In particular, in order to use
pattern recognition algorithms such as those discussed here to predict
the most damaging earthquakes, one must be able to use catalogs taken from
different fault regions for algorithm selection and optimization.  Statistics
for several large event cycles are needed to gain reliable information about
the conditional probability distributions of precursor function values which
may indicate an imminent rupture.  Recent local catalogs usually contain
only one large
event so that the utility of combining many short catalogs in an attempt to
reproduce the longer term statistics becomes an important question.

In employing algorithm M8,
Keilis-Borok and Kossobokov
make the assumption that several extremely short catalogs are
nearly as good as one very long catalog when it comes to
determining an algorithm for
forward prediction of large events.  There is really no way to test this
hypothesis well on real catalogs due to the lack of availability of catalogs
of sufficient length.  For the M8 algorithm Keilis-Borok and Kossobokov set the
parameters of their
algorithm (using eighteen possible precursor functions) to catalogs
covering over one hundred earthquakes of
magnitude greater than or equal to 8.0.  This fitting selects seven of the
precursors as most relevant and assigns them discrete threshold values.
Using these parameters the algorithm is then
applied to 44 additional catalogs from around the world containing
large earthquake epicenters.  A relevant concern is whether information on
even one hundred large event cycles is sufficient to reliably indicate seven
out of eighteen functions as viable precursors and assign them appropriate
threshold values.  Furthermore,
in the data fitting procedure, each function threshold
is evaluated individually,
then these values are retained when precursor functions are combined.
Thus it is useful to consider
how many large event cycles
 are needed in each contiguous catalog to select
reasonable threshold values for individual precursors, and hence assure
the robustness
of the  results when used for forward prediction.

Another way to phrase the problem
is in terms of the ergodicity of the
seismological record.  The assumption is that averaging over many different
earthquake fault realizations in space is equivalent to averaging over a
single fault for a much longer time.
For the UBK model this reduces to a
problem of comparing algorithm selection based upon many short-time catalogs
with that obtained at longer times.  Since the UBK model is deterministically
chaotic, it is not too surprising
that we observe
ergodic behavior.  The interesting question is what minimum
catalog length (in terms of large event cycle times) is needed for this
ergodic hypothesis to be useful?
If the minimum length were much longer than a single cycle time,
one could not hope to gain reliable predictability from any number of
catalogs containing less than one cycle's worth of data.

In order to examine this problem for our system, we
consider the distribution $P_T(Q,F_1)$ of values of $Q$
as a function of the threshold $F_1$
for an ensemble of $n$ catalogs as the
length $T$ of each of the catalogs is varied.
For each catalog length, we choose $F_1$
in order to maximize the average
$Q$, defined to be $\overline Q_{\rm max}$, and then for that
value of $F_1$ we also compute the fluctuations:
$\sigma^2(Q)=(1/n)\sum_{i=1}^n(Q_{i}^2-\overline Q_{\rm max}^2)$.
In order for optimization of the algorithm over many short
catalogs to be a useful procedure one must have available catalogs which are
sufficiently long  that, at a minimum,  $\overline Q_{\rm max}$
attains an acceptable
value and, further, the width $\sigma(Q)$ of the  distribution is relatively
small.  Otherwise, there would be no reason to believe that an algorithm
developed, for instance, using current California earthquake catalogs would
perform similarly in an independent test either on other catalogs or
for forward prediction.
Here, for convenience, we will set the spatial $\Delta s$ window
to coincide with the optimal value determined earlier in this section while
we take $\Delta t/\overline T=.18$, which allows for greater range in the
catalog lengths
which can be compared.
We compute the distribution $P_T(Q,F_1)$
by taking a single extremely long catalog (1848 large
event cycles) and breaking it down into groups of sequential catalogs of
shorter
length ranging from the time window size to several large event cycles.

Figure 15 illustrates our results for the maximized
mean value  $\overline Q_{\rm max}$, as well as
the standard deviation
$\sigma(Q)$ as a function of  catalog length $T$ normalized by the mean
large event cycle time $\overline T$.
The behavior of $\overline Q_{\rm max} (T/\overline T)$
indicates that predictability
on the UBK model is poor overall until each short catalog on average
encompasses one large event cycle. At this point $\sigma (Q)$ appears to
decrease much less rapidly as well.  Only minimal improvements are obtained
beyond a single large event cycle time (note that because the fault
is sufficiently long that large events do not encompass the entire
system, a catalog which extends for one cycle time will typically
contain several large events).

For this single predictor, the optimization after
one cycle time turns out to be a relatively simple one.
In order for a longer catalog on an individual fault to
provide  information which is not available from combining many
shorter time records, the algorithm must  make use
of correlations which develop on longer time scales.
The fact that the fluctuations in $Q$ for the activity
precursor die down on the time scale of
a single cycle implies that the activity measure does not contain appreciable
information which  is relevant on time scales longer than the time scale
over which activity builds prior to a large event.
Hence, one large event cycle is as good as another for optimization with
respect
to activity alone and this leads to the rapid decrease in fluctuations in
$\overline Q_{\rm max}$
once most of the catalogs contain at least one large event cycle.
We suspect it
is the same for the other activity-based precursor functions
considered in Section V. If such relatively short time  correlations
are dominant for seismicity on real faults,
then it is not too surprising that
the  pattern recognition algorithms such as M8
may lead to some measurable
enhancement over random prediction and  long term techniques.
On the other hand, a
full cycle's worth of data  still corresponds
to a much longer catalog than is  available
for most regions of the earth.
Thus an improvement in the stability of
algorithms may still occur as the catalogs lengthen--
especially  in regions with long repeat times.

In addition to ergodicity, the M8 algorithm assumes self-similarity
between faults in different regions
so that adjusting for the overall seismicity in an area should be sufficient
to allow transfer of an algorithm from one fault to another.  The question
of self-similarity
relates most closely to the variation of model parameters.
While we have not yet addressed this issue in great detail,
as stated previously it appears that
optimization is primarily sensitive to the UBK model parameters $\ell$ and
$\alpha$  through their determination of the length scale $\tilde \xi$
and the shape of the main events peak (for $\alpha$ and $\ell$ sufficiently
large).  These can be adjusted for by variation of the time and space windows
utilized for prediction, and hence we expect will
present no great obstacle to
combining UBK model catalogs once these are set appropriately.

\beginsection{VII. Conclusions}

We have shown how
seismicity catalogs
generated from a simple deterministic
model of an earthquake fault can provide insights
into the problem of prediction.
The ample statistics available allow for a
thorough study of both long and intermediate term  methods.
We find that while long term prediction based on recurrence intervals
performs reasonably well on the time scales that are relevant for
long term assessments, these methods are intrinsically
limited in effectiveness by the breadth of the distribution of recurrence
times. In comparison, intermediate term prediction techniques
analogous to algorithms CN and M8 do perform  better
(especially when the success curve is taken to be the measure of
performance) even when they are limited to individual
activity based precursor functions. In addition, by
optimizing with respect to algorithm parameters, we  find that
these algorithms do recognize certain fundamental
length and time scales in the UBK model. Perhaps most
importantly, at least for the simple activity precursor function,
the algorithm is as adequately optimized on many relatively short
catalogs (in which the length is approximately equal to the mean repeat
time)
as it is on one longer
catalog
(containing over 1800 large events).
While  current catalogs rarely contain a full cycle's worth of data,
it is interesting to note that one cycle time rather than many
sets the scale for the stability of the algorithm for forward
prediction.
This result suggests that predictions based on algorithms
such as CN and M8 may improve over time as the available  catalogs
become longer, and while, based on the results of the model,
the time scale over which improvements
might be expected
is certainly appreciable (of order the seismic cycle) our results
do suggest that one need
not wait forever to obtain substantial gains.

However, on a more pessimistic note,
using the most standard measure of seismicity-- the overall activity
(with or without  restricted magnitude windows)--
the intermediate
term prediction algorithms studied here
do not lead to predictions which are as precise as we had hoped.
In particular, relative to the cycle time, the alarm time is still
much too large: of order 10-20\% rather than the 1-5\% which is desired.
Perhaps these results will improve when multiple precursors are
considered. In addition, we did find that one particular measure--
the active zone size-- did perform measurably better than the rest,
capturing nearly all of the large events with roughly a 5\% alarm time.
However, it is not clear whether any such seismicity based precursor
(or combinations of precursors) can perform this well in the earth.
Ultimately  the kinds of
precursors
functions which are used in the pattern recognition algorithms
may be too primitive to  detect correlations on time scales which are
are desired for relatively precise intermediate term alerts.

Below we summarize additional aspects of the results presented in this paper.

\noindent
{\it The quality function Q:}

One methodological question we have addressed is the evaluation of the
performance of a prediction algorithm.  For this purpose we have employed
two closely related functions $Q_{P}$ and $Q_R$ which exhibit qualitatively
similar behavior.  The function $Q_P$ is a linear combination of probabilities
relevant to the prediction problem (reflecting the average success rate of
prediction, typical alarm times, and the false alarm rate) and was used here
to examine optimization questions in some detail.  The function $Q_R$, on the
other hand, is reminiscent of those traditionally employed in cost-benefit
analysis and has particularly transparent properties which enable one to
determine the public utility of any algorithm.  Both $Q_P$ and $Q_R$ measure
the extent to which  an algorithm is able to
fulfill all of the prediction goals ($Q=1$)
relative to the option of doing nothing at all ($Q=0$).
This  rather simple
specification is one which we believe will allow organizations such as
public policy
committees to more easily determine when it will be in their interest to
utilize a prediction algorithm for declaring earthquake alerts.

\noindent
{\it Long term prediction:}

We  evaluated long term predictability
for three commonly used methods:  the time-predictable and
slip-predictable models as well as
prediction based upon recurrence intervals.  The
results for the time-predictable and slip-predictable models show that neither
describes the UBK fault behavior adequately, although the time-predictable case
does appear slightly better.  It is interesting to note that
for either model
it is
possible to find good correlation over just a few subsequent large events.
At any point, however, such a short-time ``pattern'' could be broken, with the
next event differing dramatically from the proposed model.
Hence these sorts of models are certainly not sufficient when used alone
for prediction on the UBK model fault.
This is  significant for time-predictable models applied to real faults
where positive correlation is often based on knowledge of only two or three
events in a given fault region.

Prediction
based upon recurrence intervals of large events was examined using the $Q$
function.   At all times this strategy is
better than doing nothing because there are no false alarms in the way we
have chosen to implement it for the UBK model.  Clearly this is not the case
for the earth, where, for instance, the more complicated geometry allows
the accumulated strain to be relieved along neighboring interacting faults
which may not have previously hosted such a large event.
The distribution of
recurrence times is found to be fit reasonably well by a Gaussian,
with comparable results using a Weibull distribution. While neither
provides an exact fit, they
are both significantly better than a lognormal distribution.  This is due to
the greater weight at short times seen in the UBK
model.  In each fit
the standard deviation is found to be of order the mean repeat time
and the ratio of $\sigma/\overline T\sim 0.36$, which lies  between
the ratio of
$\sigma/\overline T\sim 0.75$
obtained by {\it Ward} [1992] for a different model
and  the intrinsic spread of
$\sigma/\overline T\sim 0.21$
proposed
by
{\it Nishenko and Buland}, [1987]
based on sparse
data from real faults.
In fact our results are comparable to the value of  $\sigma/\overline T
\sim .4$ which was
ultimately used
in the {\ it WGCEP}, [1988,1990] forecasts, though there the
amount in excess of $.21$ was ascribed to limited observations
rather than intrinsic spread as in the case of the UBK model.

\noindent
{\it Intermediate term prediction:}

Intermediate term precursor functions were  studied using a pattern
recognition technique  similar to those introduced and studied
by {\it Keilis--Borok et al.}, [1990{\it a, b}].
As indicated above, these methods do outperform  long term
prediction based on recurrence intervals.
When the results for different intermediate term precursors
are compared using the
$Q$ functions it is found that $Q>0$ when active zone size, activity, and
activity rate
of change are each employed as individual precursors.
Not
surprisingly, individually precursors are much more effective at
predicting events on the
UBK model than in the earth, and, in particular,
the active zone size precursor, which in
the earth is analogous to the extent to which
seismicity is broadly distributed in space,
performs remarkably well.
For this function, as well as activity and rate of change of activity,
the results we obtain
are comparable  and sometimes better, both in terms of the success curve
and the value of \qmax, than those obtained using the M8
algorithm which employs multiple precursors on real seismicity data.

However, on the UBK  model fault, activity is
a fairly direct indicator of the approach to the
instability which generates the
large events, so we had originally
expected even better quantitative results.
In fact, although we are in the process of examining functions of multiple
precursors,
it is not obvious that these changes will improve the results substantially.
While we suspect that because of its simplicity and the rise
in activity which clearly precedes almost every large
event, the UBK model will
ultimately be more
amenable to prediction than will the earth,
this need not  $a$ $priori$ be so.
Because  of the breadth of the distribution of activity
values prior to a large event, we suspect
some other feature will be necessary
in order to make predictions  with alarm times
which are significantly less.
Thus
it may ultimately be that the heterogeneity of the earth will lead to
phenomena which probe a wider degree of time scales than those which are
available in the UBK model, and  ultimately
then also to more precise predictions.

\noindent
{\it Algorithm Optimization:}

In the UBK model,
optimization of the $Q$ function for activity with respect to space and
time windows leads to
selection of lengths which scale with  $\tilde \xi$ and time windows of order
$1/3$ the mean recurrence interval, which are the length and time scales
over which correlations have been previously observed to grow prior to a large
 event [{\it Carlson and Langer}, 1989{\it b}; {\it Shaw et al.}, 1991].
It is a positive feature that the algorithms recognize these relevant length
and time scales, however, it leaves open the question of what will happen
when the algorithms are applied to fault models for which there is no
break in the scaling behavior.  In such models (see, e.g.
{\it Chen, et al.} [1991]) typically
the dynamics have no inherent
length or time scale so the behavior of $Q$ as a function of box size
may be very different.

The optimization issue perhaps most relevant to applications of the pattern
recognition algorithms to the earth,
is that of stability of prediction for independent data (catalogs not used
during the learning procedure).  In the model, we find that for activity
alone, catalogs containing one large event cycle are sufficient for stability.
If in the earth correlations in seismicity based precursors
also do not extend  in time beyond a single cycle,
then  our results suggest that
substantial improvements in predictability
may be expected as data bases
increase to times of order the mean repeat time for large events.

\noindent
$Acknowledgements$:
In the course of this  work we have profited greatly from
discussions with
V.~Keilis-Borok, A.~Gabrielov, D.~Turcotte, J.~Dieterich,
G.~Swindle, and
especially
J.~S.~Langer.
The work of JMC was supported by a grant from the Alfred P.
Sloan Foundation and NSF grant DMR-9212396.
The work of SLP was supported by an INCOR grant from the CNLS at
Los Alamos National Laboratories.
The work of BES was supported by the SCEC grant USC-572726,
and USGS grant 1434-93-G-2284.
The work of JMC and BES  was also supported  by
the National Science Foundation under Grant
PHY89-04035.

\centerline {REFERENCES}

\noindent
Aki, K.J., Magnitude-frequency relation for small earthquakes; a clue to the
origin of $f_{max}$ of large earthquakes, {\it J. Geophys. Res., 92},
1349-1355 (1987).

\noindent
Archuletta, R.J., E. Cranswick, C. Mueller, and P. Spudich, Source parameters
of the 1980 Mammoth Lakes, California earthquake sequence,
{\it J. Geophys. Res., 87},  4595-4607,
1982.

\noindent
Bak, P., C. Tang, and K. Wiesenfeld, Self-organized criticality:  an
explanation of 1/f noise, {\it Phys. Rev. Lett., 59}, 381-384, 1987.

\noindent
Bakun, W.H., C.G. Buffe, and R.M. Stewart,
Body wave spectra of central California earthquakes,
{\it Bull. Seismol. Soc. Am., 66}, 363-84, 1976,

\noindent
Burridge, R., and L. Knopoff, Model and theoretical seismicity, {\it Bull.
Seismol. Soc. Am., 57}, 3411-3471, 1967.

\noindent
Carlson, J.M., Time intervals between characteristic earthquakes and
correlations with smaller events:  an analysis based on a mechanical model of
a fault, {\it J. Geophys. Res., 96}, 4255-4267, 1991.

\noindent
Carlson, J.M., J.S. Langer, B.E. Shaw, and C. Tang, Intrinsic properties of
a Burridge-Knopoff model of an earthquake fault, {\it Phys. Rev. A, 44},
884-897, 1991.

\noindent
Carlson, J.M. and J.S. Langer, Properties of earthquakes generated by fault
dynamics, {\it Phys. Rev. Lett., 62}, 2632-2635, 1989{\it a}.

\noindent
Carlson, J.M. and J.S. Langer, Mechanical model of an earthquake
 fault, {\it Phys. Rev. A, 40}, 6470, 1989{\it b}.

\noindent
Chen, K., P. Bak, and S. Obukov, Self-organized criticality
in a crack-propagation
model of earthquakes, {\it Phys. Rev. A 43}, 625-630, 1991.

\noindent
Davis, P.M., D.D. Jackson, and Y.Y. Kagan, The longer it has been since the
last earthquake, the longer the expected time till the next?, {\it
Bull. Seismol. Soc. Am., 79}, 1439-1456, 1989.


\noindent
Davison, F.C., Jr., and C.H. Scholz, Frequency-moment distribution of
earthquakes in the Aleutian Arc:  a test of the characteristic earthquake model
, {\it Bull. Seismol. Soc. Am., 75}, 1349-1362, 1985.

\noindent
Dieterich, J.H. An alternate null hypothesis,
{\it U.S. Geol.
Sur. Open File Rep. 88-398}, 1992.

\noindent
Gabrielov, A.M., T.A. Levshina, and I.M. Rotwain, Block model of earthquake
sequence, {\it Phys. Earth Planet. Inter., 61}, 18-28, 1990.

\noindent
Healy, J.H., V.G. Kossobokov, and J.W. Dewey,
A test to evaluate the earthquake prediction algorithm M8,
{\it U. S. Geo. Sur. Open File Rep.  92-401}, 1992.



\noindent
Kanamori, H., The nature of seismicity patterns before large earthquakes,
in
{\it Earthquake Prediction, an International Review},
Maurice Ewing Series, Vol. 4, ed. D.W. Simpson and P.G.Richards,
Amer. Geo. Un., Washington, D.C., 1-19, 1981.

\noindent
Keilis-Borok, V.I., and I.M. Rotwain,
Diagnosis of time of
increased probability of strong earthquakes in different regions of the world:
algorithm CN, {\it Phys. Earth Planet. Inter., 61}, 57-72, 1990{\it a}.

\noindent
Keilis-Borok, V.I., and V.G. Kossobokov, Premonitory activation of
earthquake flow: algorithm M8, {\it Phys. Earth Planet. Inter., 61},
73-83, 1990{\it b}.

\noindent
Keilis-Borok, V.I., L. Knopoff, V.G. Kossobokov, and I. Rotwain, Intermediate-
term prediction in advance of the Loma Prieta  earthquake, {\it Geophys. Res.
Lett., 17}, 1461-1464, 1990c.

\noindent
Kiremidjian, A.S. and T. Anagnos, Stochastic slip-predictable model for
earthquake occurrences, {\it Bull. Seismol. Soc. Am., 74}, 739-755, 1984.

\noindent
Langer, J.S. and C. Tang, Rupture propagation in a model of an earthquake
fault, {\it Phys. Rev. Lett., 67}, 1043-1046, 1991.

\noindent
Langer, J.S., Models of crack propagation, {\it Phys. Rev. A 46}, 3123
(1992).

\noindent
Malin, P.E., S.N. Blakeslee, M.G. Alvarez, and A.J. Martin, Microearthquake
imaging of the Parkfield asperity, {\it Science, 2 44},
 557, 1989.

\noindent
McNalley, K.C., and J.B. Minster, Nonuniform seismic slip rates
along the Middle American Trench, {\it J. Geophys. Res., 86}, 4949-4959
1981.

\noindent
Mogi, K., Study of elastic shocks caused by the fracture of heterogeneous
material and its relation to earthquake phenomena, {\it Bull. Earthq. Res.
Inst., Univ. Tokyo, 40}, 125-173, 1962.

\noindent
Molchan, G.M., and Kagan, Y.Y., Earthquake prediction and its optimization,
{\it J. Geophys. Res., 97}, 4823-4838, 1992.

\noindent
Molchan, G.M., Structure of optimal strategies in earthquake prediction,
{\it Tectonophysics, 193}, 267-276, 1991.

\noindent
Nishenko, S.P., and R. Buland, A generic recurrence interval distribution
for earthquake forecasting, {\it Bull. Seismol. Soc. Am., 77}, 1382-1399, 1987.

\noindent
Pacheco, J.F., C.H. Scholz, and L.R. Sykes, Changes in frequency-size
relationship from small to large earthquakes, {\it Nature, 235}, 71-73, 1992.

\noindent
Scholz, C.H., Earthquake prediction and seismic hazard, {\it Earthq. Predict.
Res., 3}, 11-23, 1985.

\noindent
Schwartz, D.P. and K.J. Coppersmith, Fault behavior and characteristic
earthquakes: examples from the Wasatch and San Andreas fault zones, {\it J.
Geophys. Res., 89}, 5681-5698, 1984.

\noindent
Shaw, B.E., J.M. Carlson, and J.S. Langer, Patterns of seismic activity
preceding large earthquakes, {\it J. Geophys. Res., 97},
 479, 1992.

\noindent
Shaw, B.E., Moment spectra in a simple model of an earthquake fault,
{\it Geophys. Res. Lett.}, to appear, 1993 {\it a}.

\noindent
Shaw, B.E., Generalized Omori Law for aftershocks and foreshocks
from a simple dynamics,  {\it Geophys. Res. Lett.}, to appear, 1993 {\it b}.

\noindent
Shimazaki, K. and T. Nakata, Time predictable recurrence model for large
earthquakes, {\it Geophys. Res. Lett., 7}, 279-282, 1980.

\noindent
Sykes, L.R. and M. Tuttle, A seismic precursor to the Loma Prieta earthquake,
rupture zones of historic nearby earthquakes and seismic potential of the
Hayward fault, {\it USGS Monograph}, March 25, 1991.

\noindent
Thatcher, W., The earthquake deformation cycle, recurrence, and the time-
predictable model, {\it J. Geophys. Res., 89}, 5674-5680, 1984.

\noindent
Vasconcelos, G.L., M.S. Vieira, and S.R. Nagel, Phase transitions in a spring-
block model of earthquakes, {\it Physica A, 191}, 69-74, 1992.

\noindent
Ward, S.N., An application of synthetic seismicity in earthquake statistics -
the Middle America Trench, {\it J. Geophys. Res., 97}, 6675-6682, 1992.

\noindent
Wesnousky, S., C.H. Scholz, K. Shimazaki, and T. Matsuda, Earthquake frequency
distribution and the mechanics of faulting, {\it J. Geophys. Res., 88},
9331-9340, 1983.

\noindent
Working Group on California Earthquake Prediction,
Probabilities of large earthquakes occurring in California on the San
Andreas fault, {\it U.S. Geol.
Sur.  Open File Rep. 88-398}, 1988.

\noindent
Working Group on California Earthquake Prediction,
Probabilities of large earthquakes in the San Francisco
Bay Region, California, {\it U.S. Geological
Survey Circular, 1053}, 1990.

\noindent
Wyss, M., Precursors to large earthquakes, {\it Earthq. Predict. Res., 3},
519-543, 1985.

\endreferences
\vfill\eject

\centerline{FIGURES}
\doublespace

\noindent
(1). A sample catalog illustrating the events which take place as a
function of space and time
in the UBK model.
Time on the vertical axis is measured  relative to  the inverse loading
speed,
so that values on that axis represent  $t\nu=\delta U$, i.e.~the
net  displacement of initially adjacent points
on opposite sides of the fault.
A line segment  is drawn through all of the
blocks which slip during an event, and a cross marks the position of the
epicenter of each large event.
Individual events are also quite complex, with spatially irregular slip
which unfortunately
cannot be illustrated simultaneously on a plot such
as this.
Figure (a) represents only a fraction (both in space and time) of the
full catalog  which is
considered, and (b) is an expansion of the precursory activity
for one the large events in (a).
Unless explicitly  stated
otherwise,  the numerical results  illustrated
in the figures that follow will be for
a system of size $N=8192$, with $\ell=10$, $\sigma=.01$ and
$\alpha=3$. The full catalog (excluding one block events-- the exclusion
of these was not seen to alter the results) consists of
114,000 events, 1848 of which are large events and corresponds to a total
displacement of $\delta U=122$.

\noindent
(2). Log Frequency ln[$D(\mu)$] vs.~magnitude $\mu$ for the UBK model.
Here $D(\mu)d\mu$ is  the  number of events per unit length per unit time
(time is measured in units of the inverse loading speed, as in Fig.~1)
in the magnitude range $[\mu,\mu+d\mu]$.
The small events satisfy the Gutenberg--Richter law
$D(\mu)=Ae^{-b \mu}$ with $b=1$, while the large events
occur at a rate which is in excess of the extrapolated rate of small
events.
The crossover between small and large events is denoted by $\tilde {\mu}=
\rm {ln} \widetilde M$, where $\widetilde M \approx 2 / \alpha$.
The peak in the large events distribution corresponds roughly to a rupture
length $\xi^*\approx 10\tilde\xi$ which is well below the system size.
The magnitude $\mu^*$ typically associated with such events is indicated in
the figure.
In the corresponding integrated distribution (i.e.~where the
frequency of events with magnitude in the range $[\mu,\infty)$ is plotted)
the peak is replaced by a flat shoulder.
We choose to plot the differential distribution
because
the location of the
crossover magnitude $\tilde\mu$ is suppressed in the integrated distribution.

\noindent
(3). Cumulative slip as a function of time for a
representative patch  along the fault.
While the small events overall are much more frequent than the large
events,
all of the visible displacement is associated with large events.
The lines illustrate the best least-squares fits to the upper and
lower corners of the
staircase, and thus represent the best fits to the
slip-predictable and time-predictable models, respectively.
Clearly, the sequence of events is not periodic, and although
neither the slip-predictable model  nor the time-predictable
model work particularly well, a more complete test (see Fig.~4)
indicates that the time-predictable model does somewhat better
(in particular, the root mean square deviation from the linear
fit is less  for the  time-predictable model).

\noindent
(4). Tests for correlations between seismic moments and time intervals
for subsequent large events. Here
$T_i$ and $T_{i+1}$ are the time intervals
preceding large subsequent events of moment $M_i$ and $M_{i+1}$,
respectively. Pairs of events in which the epicenters lie within $\tilde\xi$
of one another are considered.
Fig.~(a) illustrates a test of the time-predictable model
which, if correct, would lead to collapse of  the data  for
$T_{i+1}$ vs.~$M_i$
onto a line.
Fig.~(b) illustrates a test of the slip-predictable model
which, if correct, would lead to a collapse of the data for
$M_{i}$ vs.~$T_i$ onto a line.
In (c) and (d) we  test for correlations between
$M_i$ and $M_{i+1}$ as well as $T_i$ and $T_{i+1}$, respectively.
While none of the graphs show significant correlations,
compared to the others
(a) minimizes the least squares deviation to the best linear fit, which
is shown in that case.
However, our primary conclusion from this data is that while
short catalogs may sometimes suggest some correlation
such as those tested here,
none of these
simple prediction schemes  is reliable even for the simple UBK model.

\noindent
(5). Local recurrence intervals for large events on the UBK model, scaled
by the mean recurrence interval
$\overline {T}=.558/\nu$. Figure (a) illustrates the density
(i.e.~the probability of a repeat time in the interval
$[T/ \overline {T},
T/ \overline {T}+d T/ \overline {T}]$)
while (b) represents the cumulative distribution along with the best fits
to  Gaussian (G), Weibull (W), and lognormal (L) distributions (shifted
for convenience).
Comparison of the  root mean square deviation for each of the fits indicates
that the Gaussian (starting at $T/\overline T=0$, i.e.~omitting
unphysical negative
intervals) provides
the best fit, followed closely by the Weibull, and finally by the lognormal.

\noindent
(6). Success curve for time interval prediction.
Here we plot the fraction of large events predicted ($p_1$) vs.~the fraction
of time  occupied by alarms ($p_2$).
The octagons correspond to local predictions, while the dashed line
corresponds to predictions made in spatially coarse grained windows
of size $3\tilde\xi$ (we show the second curve for comparison to the
results obtained for the intermediate term algorithms).
The upper curve represents a limit
for prediction based upon recurrence intervals for the model.  In this case,
predicting $80 \%$ of the large events requires that alarms be declared
roughly $35 \%$ of the time.
Each point on the curve corresponds to a different value of the
threshold time $t_0$ at which an alarm will be issued
relative to the time of the last large event.
This curve can be deduced directly from the time interval distribution
illustrated in Fig.~5a, where, for example, the
fraction of events successfully predicted
corresponds to the  relative weight
of intervals greater than $t_0/\overline T$ in the density.

\noindent
(7). Quality function $Q$ for time interval prediction.
Here we plot $Q=p_1 - p_2$
($p_3=0$ since there are no false alarms with this method)
as a function of $p_2$ for the data
illustrated in Fig.~6.
As previously octagons represent  local forecasts, and the
dashed line represents the coarse grained results.

\noindent
(8). Distribution of values of the activity precursor $f_1(\Delta s,
\Delta t)$
evaluated in space-time windows
$\Delta s=3 \tilde {\xi}$  and
$\Delta t/\overline T=.36$.
Both the full time average $P(f_1)$ as well as the distribution
of values just prior to a large event
$P(f_1|$ large event) are illustrated.
The
relatively small  overlap between the two distributions suggests
that activity should be a reasonably good intermediate term predictor
for the UBK model.

\noindent
(9). Success curves for intermediate term
precursor functions
$f_i(\Delta s=3\tilde\xi,\Delta t/\overline T=.36)$.
Note that for  purely
random prediction the fraction predicted $p_1$ is
equal to the fraction time occupied by alarms $p_2$  (i.e.~the straight line
on the graph)
indicating that  all of the functions perform better than random prediction.
In terms of the success curve, all of the functions perform better
than prediction based on time intervals as well (i.e.~at least for some
range of $p_2$ for the success curve for time
interval prediction represented here by the dashed
line lies below each of the curves shown).
It is interesting to note
the extent to which active zone size outperforms the
other precursors, predicting nearly all of the large events
when alarms occupy  roughly 5\% of the total time.
While this is clearly the best indicator,
active zone size, activity, and activity rate of change all perform
substantially
better than random prediction
for all threshold values, predicting 90\% of events with
alarms on 5\%, 14\%, and 16\% of the time, respectively.  Fluctuations
in activity performs
somewhat worse than the others,
achieving a probability gain of
only about 2.7 over random prediction, compared to the
long term prediction gain of 2.3 (when 90\% of events are predicted).

\noindent
(10). Quality functions $Q$ for the intermediate term
precursor functions illustrated in Fig.~9.
Figure (a) represents results for the probability based function
$Q_P$ with $Q_P=p_1-p_2-p_3$, and the $p_i$ are taken to be probabilities,
while (b) represents results for the cost-benefit
function
$Q_C$ with $Q_C=p_1-p_2-p_3/3$, and the $p_i$ are taken to be rates.
In either case, when $Q<0$,
from a cost--benefit standpoint it
is better to do nothing (then $Q=0$).  Active zone size, activity, and activity
rate of change all perform reasonably well compared to doing nothing,
with
their $Q$ values  rising above zero by the time alarms occupy
$5 \%$ of the total observation period. In comparison,
fluctuations in activity
almost always has $Q<0$,  indicating that it is a rather poor measure.
Compared to the relative results for the success curve (Figs.~6 and 9)
in which all of the intermediate term precursors outperformed
the  long term technique,
here values of $Q$ are more comparable to  (and sometimes worse than)
those obtained
for  time interval prediction (Fig.~7) because of the penalty for false alarms
which  is significant for all of the intermediate term precursors
but which is
not relevant for time interval prediction.  Note that in some applications
one might be willing to tolerate a higher percentage of false alarms, and
thus choose to reduce the coefficient
$|A_3|$ of $p_3$ in $Q$.  In
that case all of the precursors considered here would have increased
values of $Q$ and thus  have $Q>0$ most of the time.

\noindent
(11). Test for independence of activity $f_1$ and rate of change of activity
$f_2$
precursor functions. Each point in  (a) represents   simultaneous
values of $f_1$ and $f_2$  at unrestricted points in time
while the values plotted in (b) are the values just prior to
a large event.
These plots constitute projections of the  joint probability distributions
for $f_1$ and $f_2$ analogous to  those illustrated for $f_1$ in Fig.~8.
While the values are clearly correlated,
the  nonnegligible spread in the joint distribution prior to a large
event suggests that the quality of our
predictions  may be enhanced through a judicious
combination of the two functions as described in the text.

\noindent
(12). Optimization of space and time windows for the activity precursor
$f_1$.
Here \qmax is plotted as a function of $\Delta s$ and $\Delta t/\overline T$
for the probability based quality function $Q_P$.
At each space and time window, $Q_P$
has already been optimized with respect to
the threshold value $F_1$.
Here the maximum with respect to $\Delta s$ and $\Delta t$
is fairly broad
indicating that the algorithm is robust  with respect to variations
in these parameters.
In particular, variations in \qmax are less than 10\% for time windows
from ranging from $.2$ to $.5$ of
the mean recurrence interval and
space windows ranging from $2\tilde\xi$ to $7\tilde \xi$.

\noindent
(13). Cross-sections for (a) space and (b) time
window optimization.
Here we illustrate the danger of an injudiciously chosen quality
function $Q$. In each case the  lower curve is simply a cross-section
of the results plotted in Fig.~12 ($ \vert A_3 \vert =1$),
while the upper curve corresponds
to the results obtained when the penalty for false alarms
is not included ($ \vert A_3 \vert =0$) . In (a) the intermediate value
$ \vert A_3 \vert =.3$ is  included
to illustrate that the size of the  optimal  spatial window is
sensitive to the choice of this coefficient.
The fact that the upper curves $(|A_3|=0)$ are maximized
for small (essentially infinitesimal)
space and time windows
leads to an ill-defined optimization problem as described in the
text.

\noindent
(14). The effects of restricted  magnitude windows. Here we plot
\qmax for the  activity precursor
as a function of lower magnitude cutoff $\mu_{\rm min}$ for the
catalog (events below the cutoff are omitted in our evaluation of
$f_1$).  The steady rise
in Q up to $\mu_{\rm min}\approx-7.5$ is due to the decreasing number of false
alarms.   The curve plateaus near where Q is optimized
for a threshold value of 1 event, and thereafter is governed by the probability
of observing a single event of sufficient magnitude to trigger an alarm before
the large event is initiated.  The steep decline occurs roughly when the
integrated
probability of observing an event above the cutoff per unit length per unit
time is less than the probability of observing a large event.
Here we show results for $\Delta t/\overline T=.18$
in order to clearly illustrate the
bump present for intermediate magnitudes, suggesting a possible additional
precursor function associated with intermediate size events.

\noindent
(15). The effects of   restricted catalog length.
Here we consider the convergence of the  intermediate term
prediction algorithm as a function  of the catalog length (in time).
The space and time windows for the algorithm
are  fixed at $\Delta s=3\tilde\xi$ and
$\Delta t/\overline T=.18$ (the optimal $\Delta t$ was
not chosen in order to allow for greater range in catalog length).
In each case, roughly 100 catalogs were  combined, and
optimized with respect to the threshold $F_1$, so that
for each catalog length
$F_1$ was set by the  maximum average
$\overline Q_{\rm max}$ for the set of catalogs of that
length.
In (a) we plot  this average value as a function of catalog length.
In (b) we plot the width of the distribution of $Q$ values
(for the independent catalogs with the threshold fixed at $F_1$)
as a function of catalog length.
Interestingly  the mean converges to its optimal value and the width
approaches zero once the  catalog length becomes roughly
of order the mean recurrence interval.

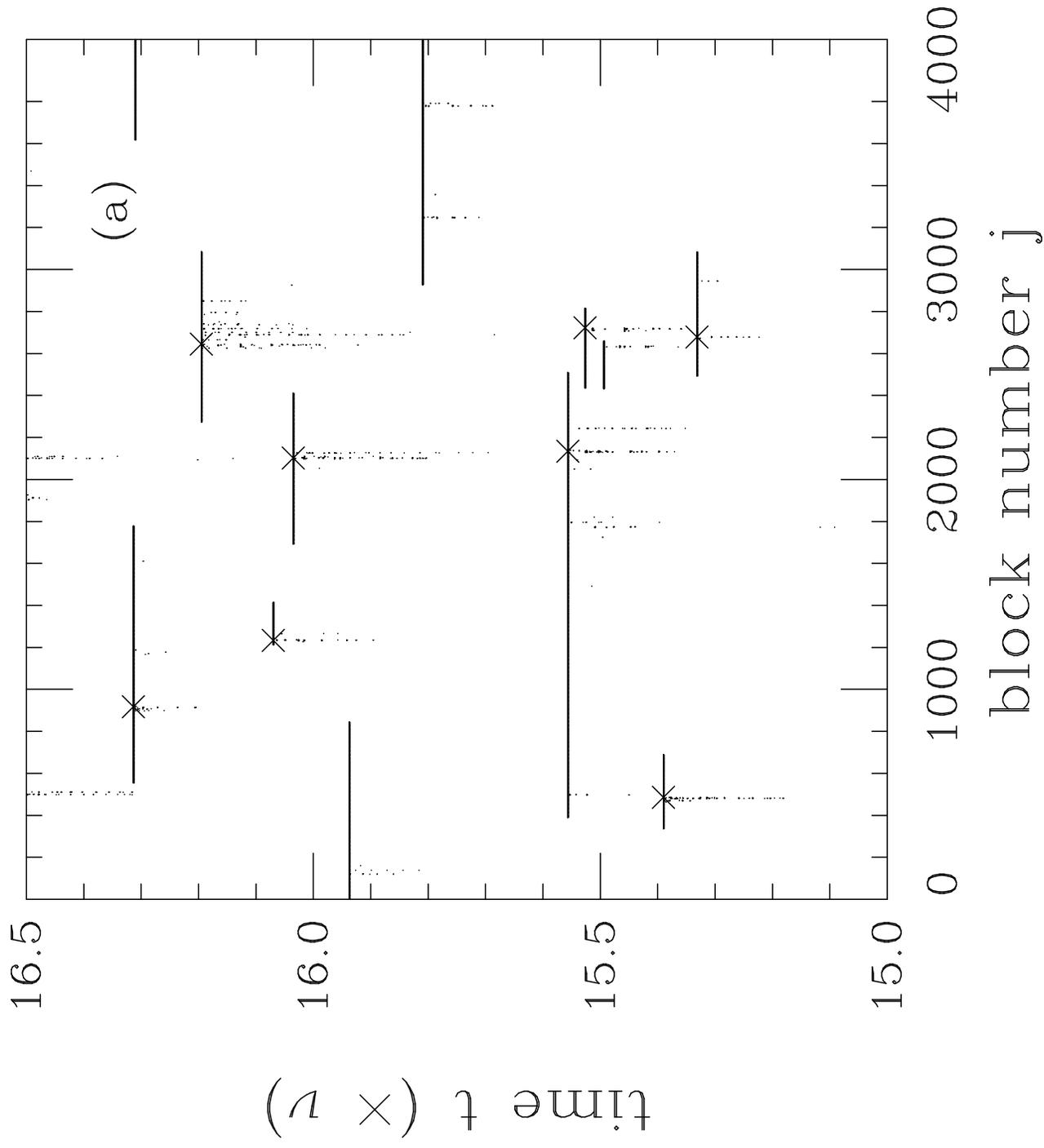
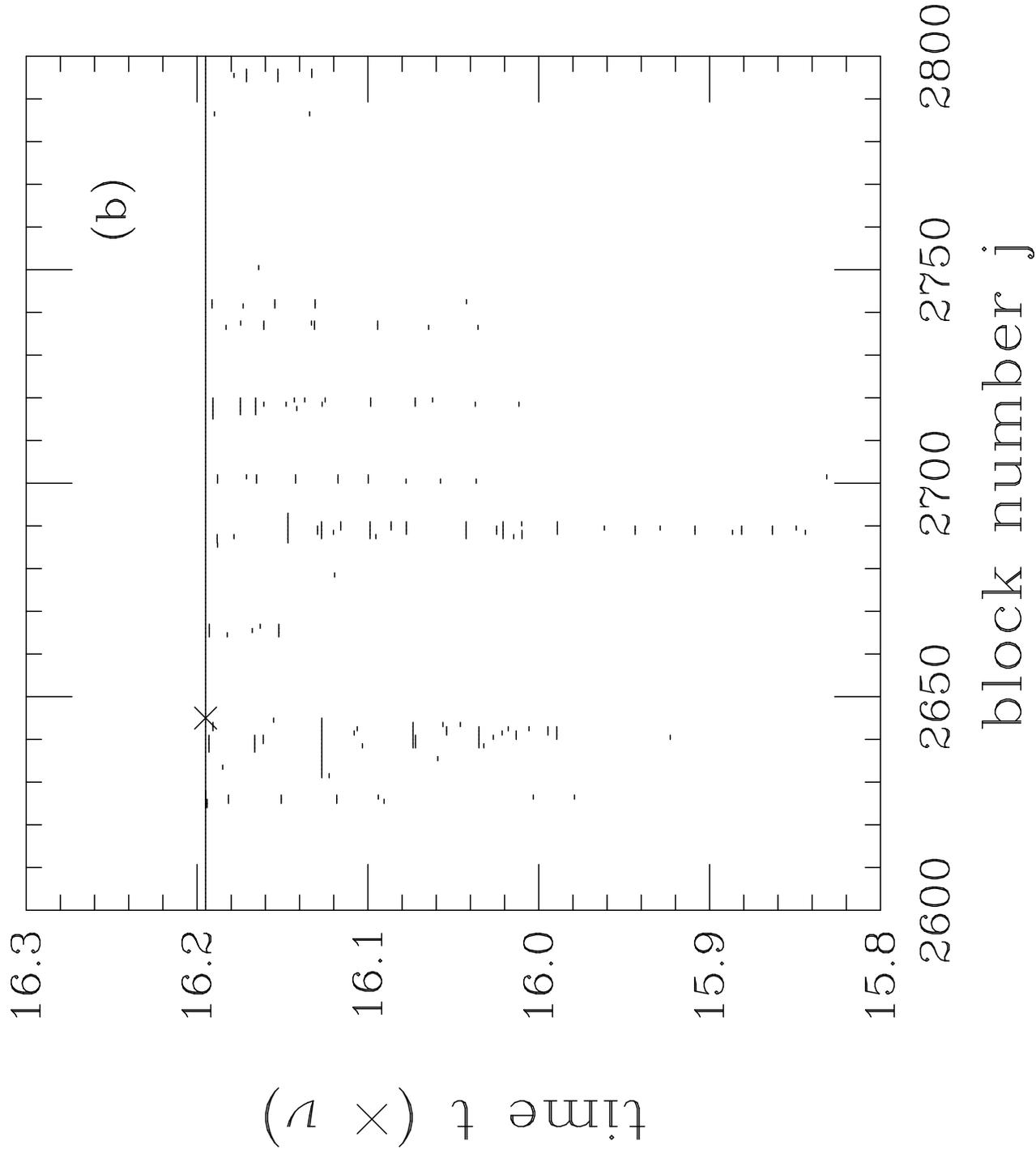
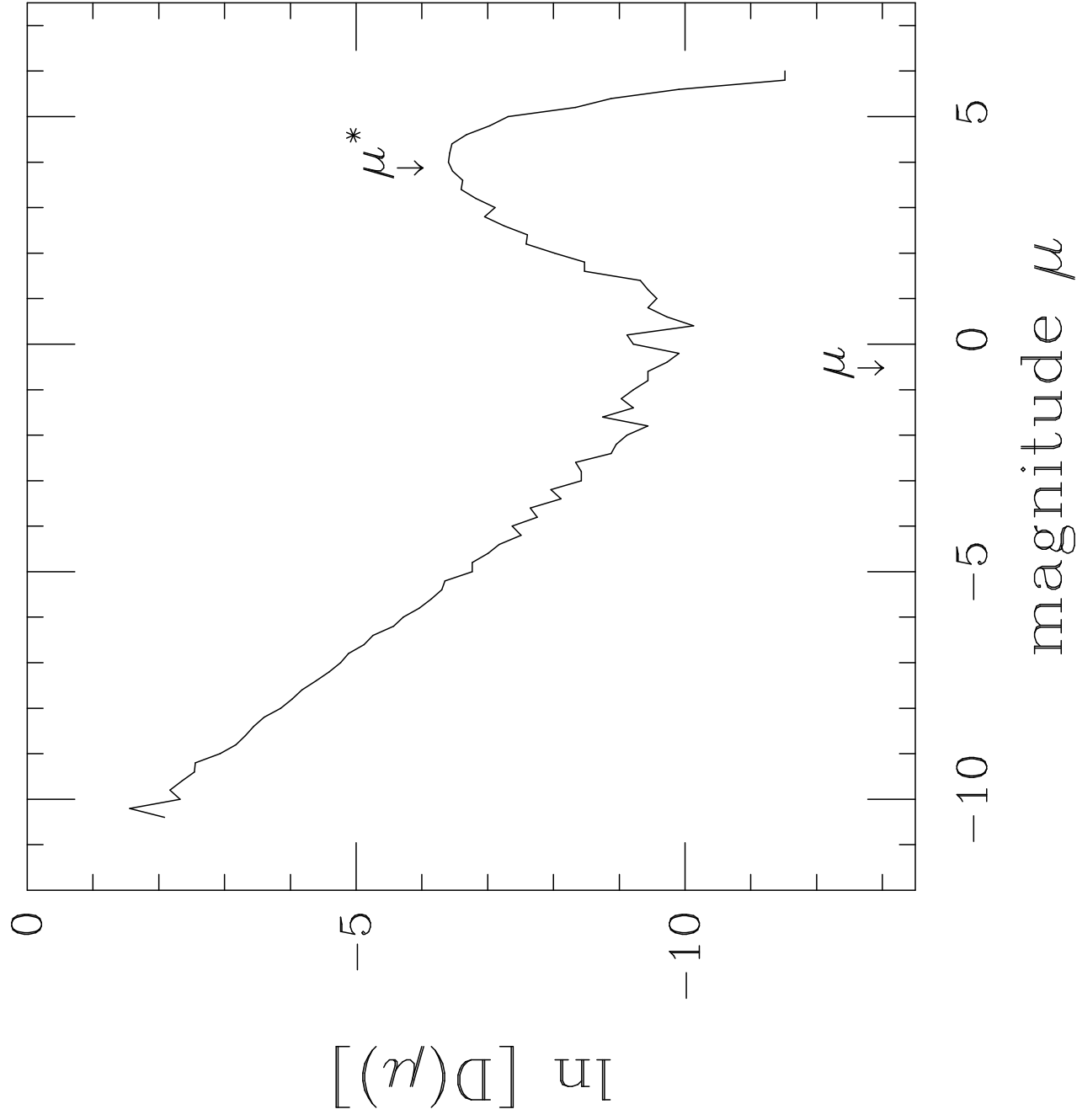
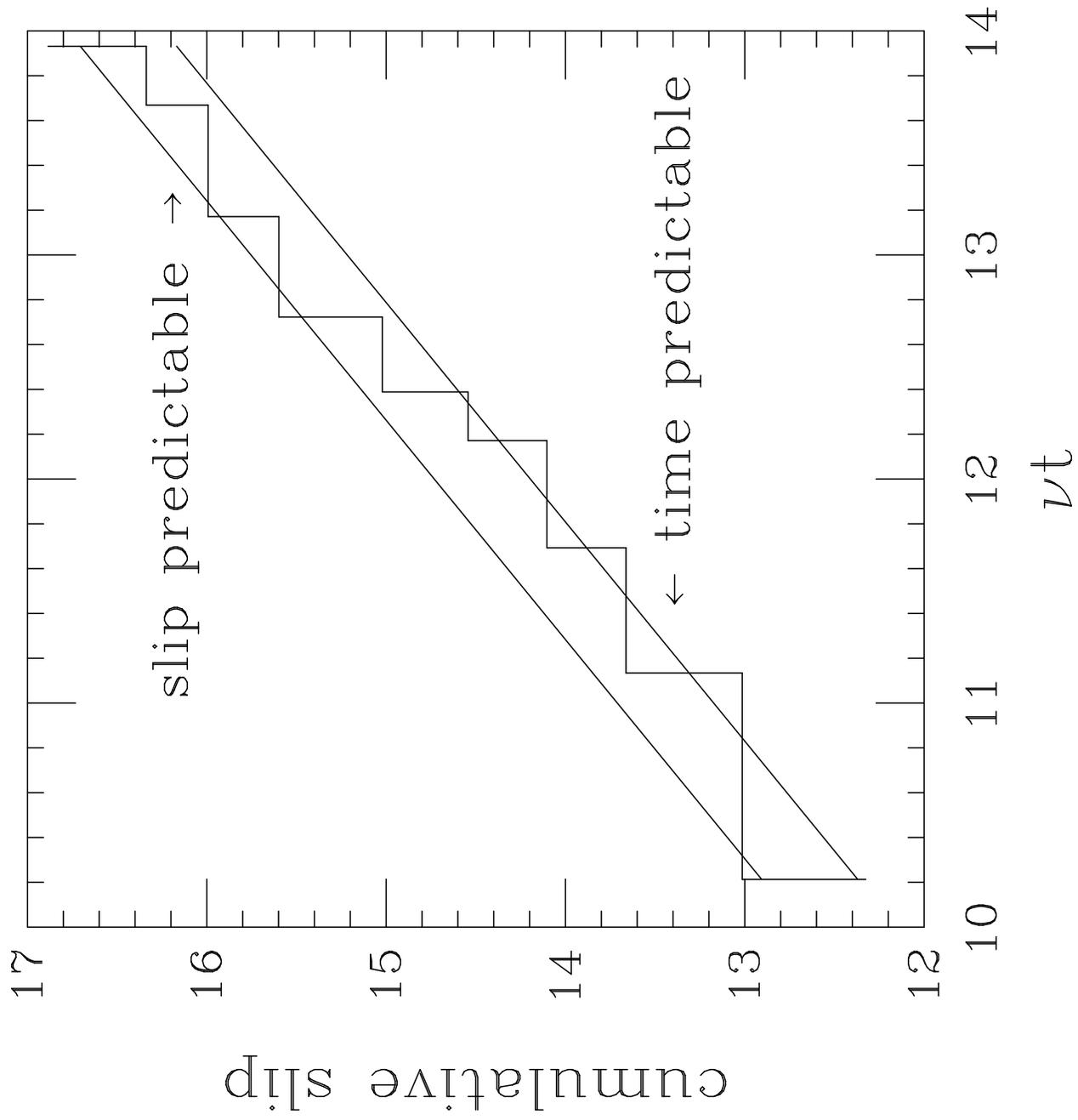
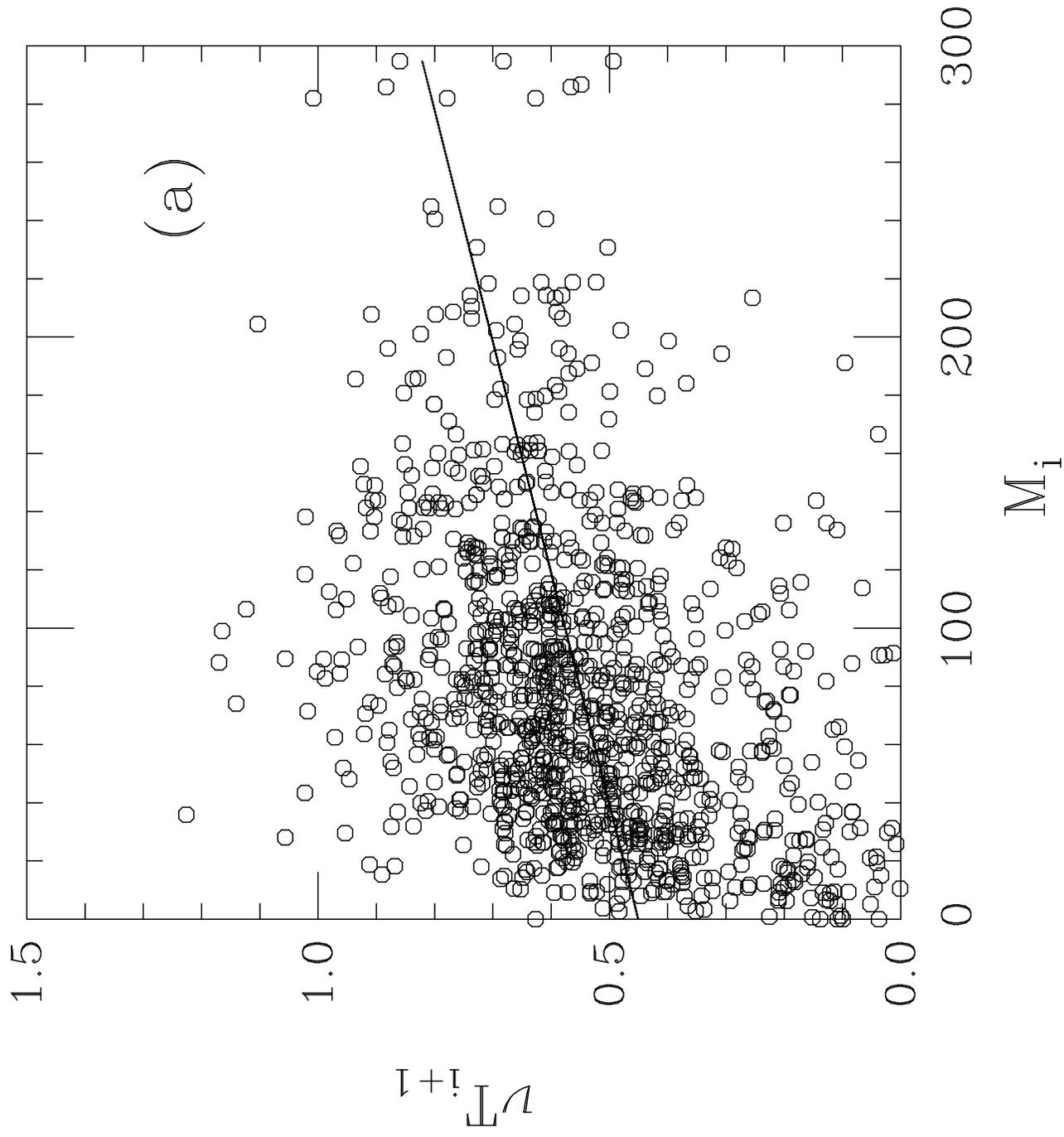
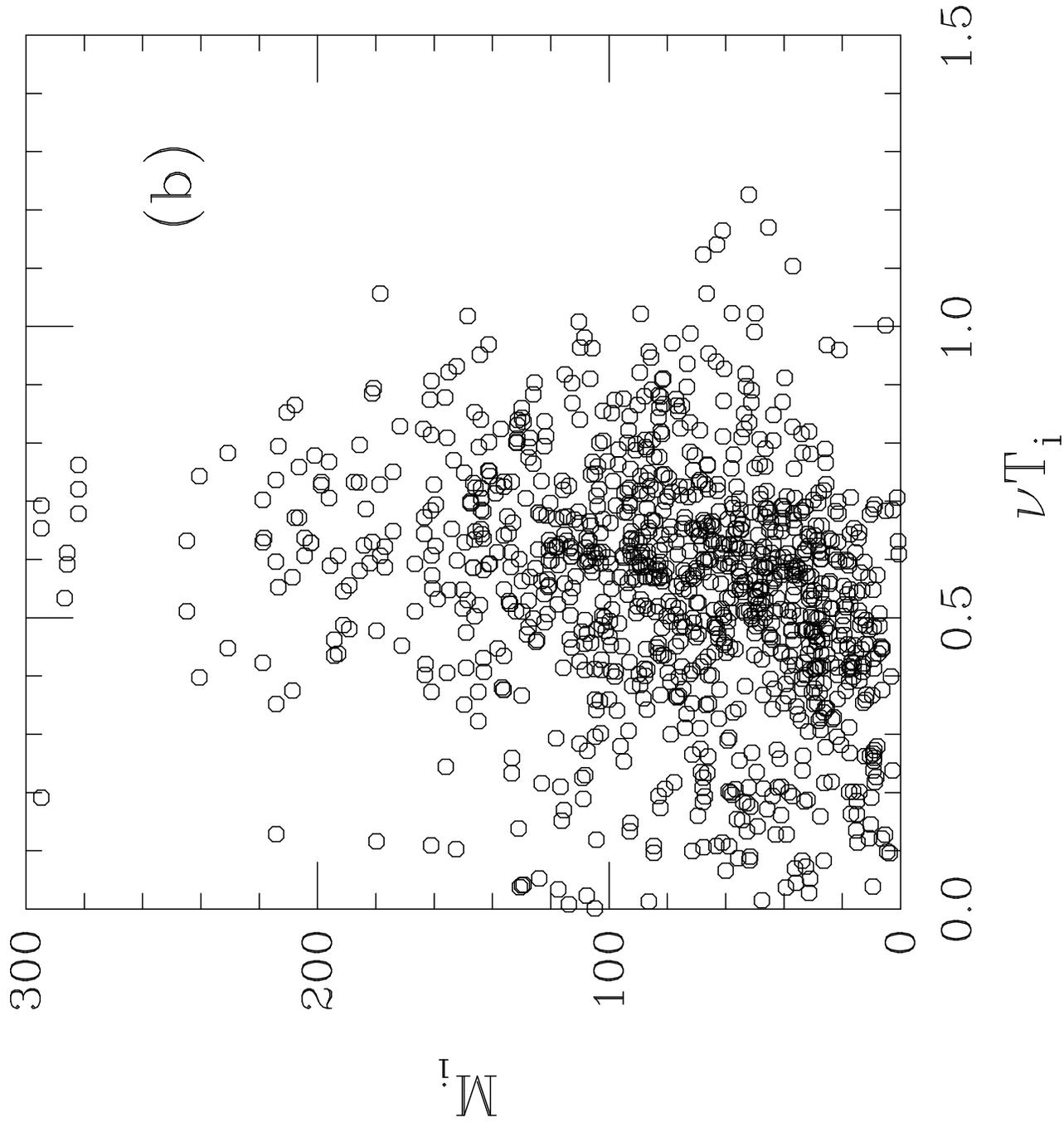
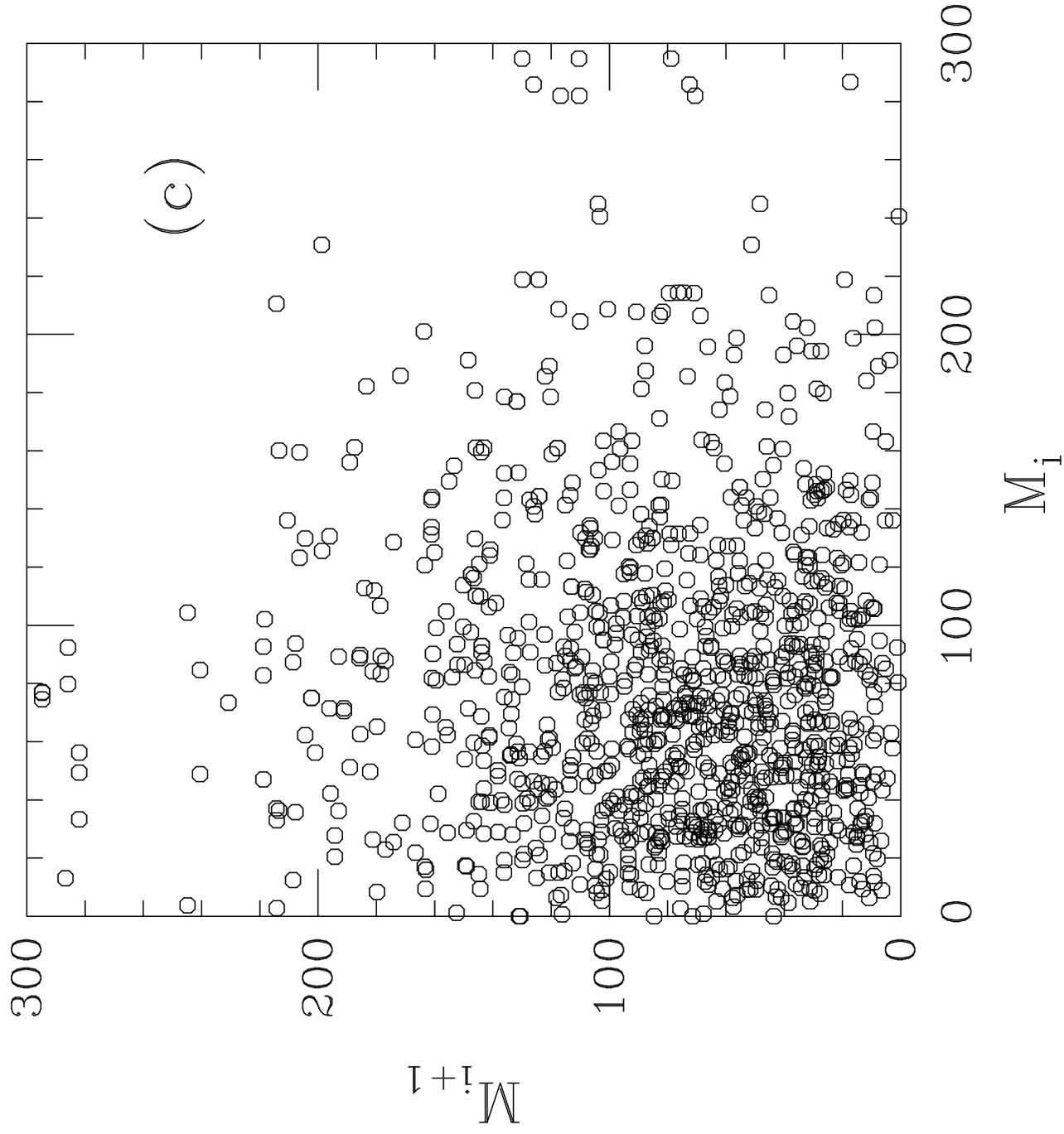
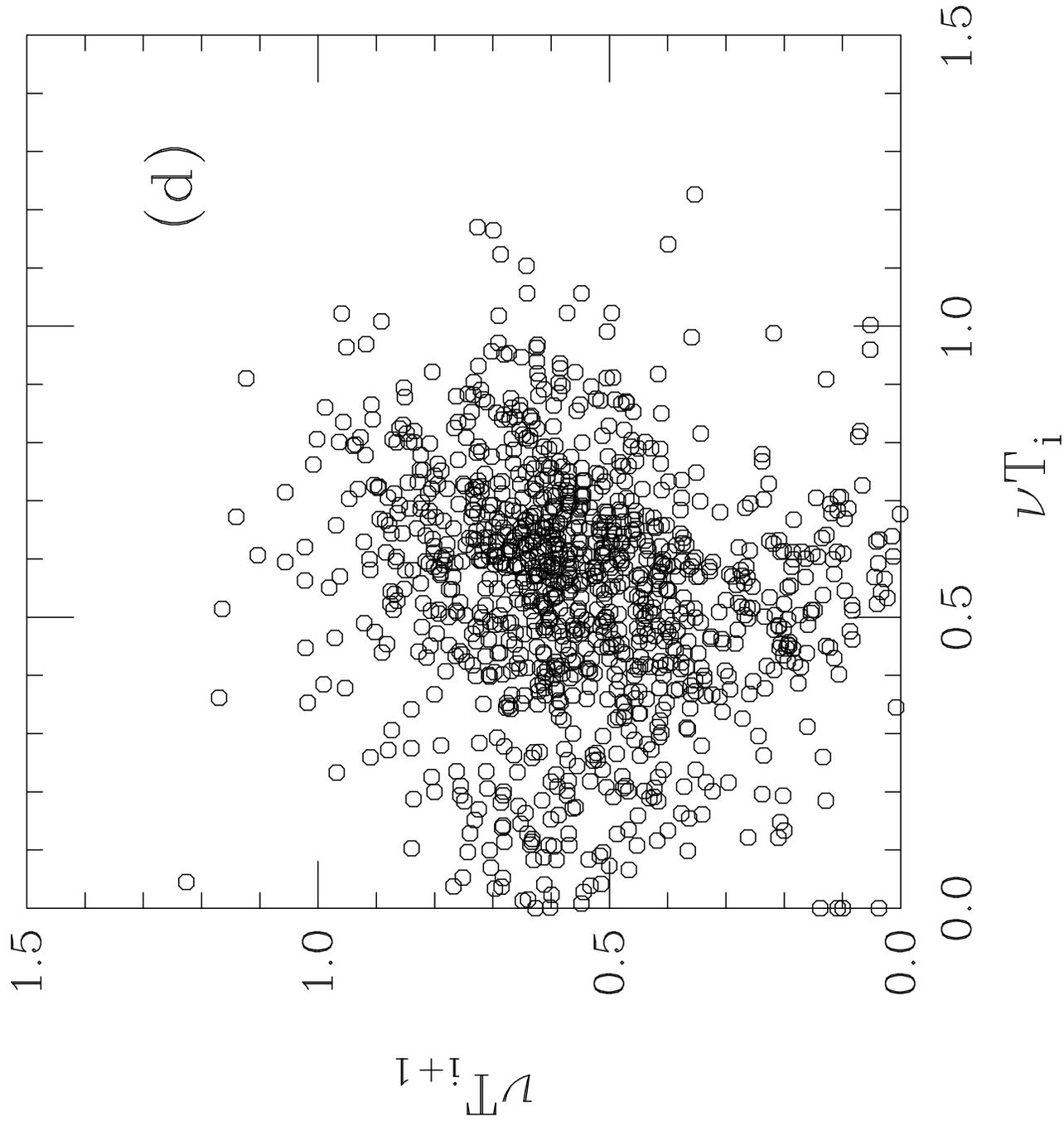
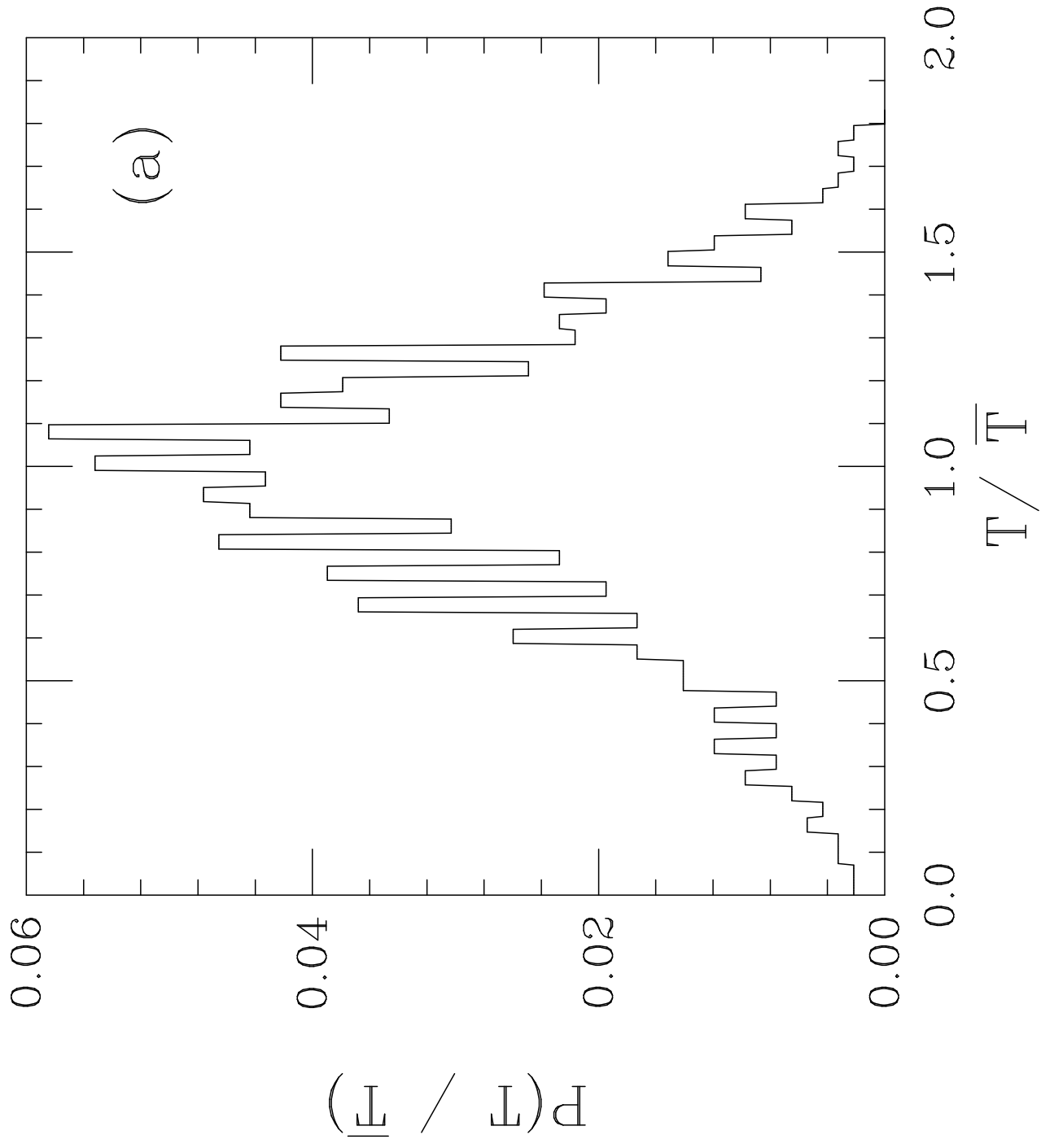
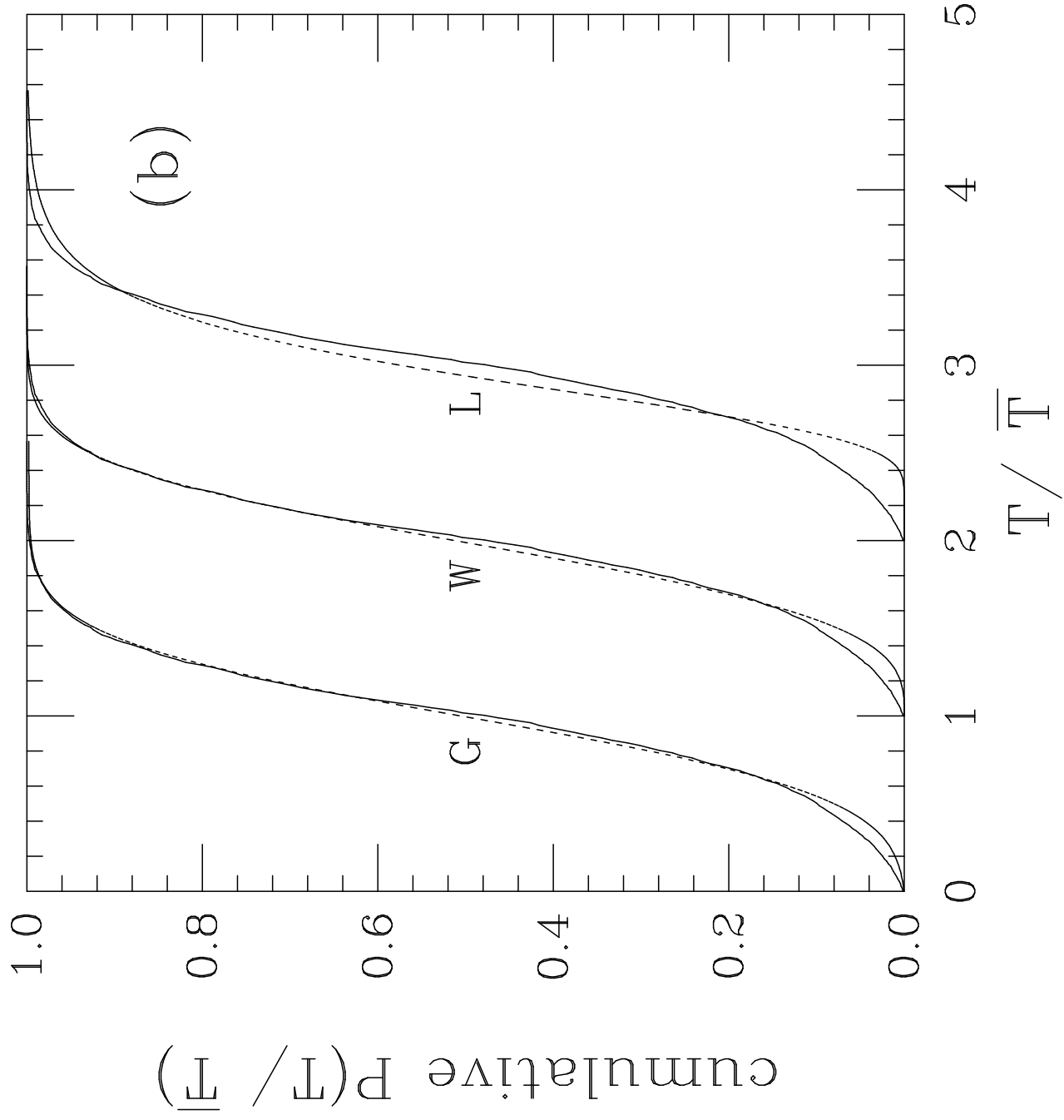
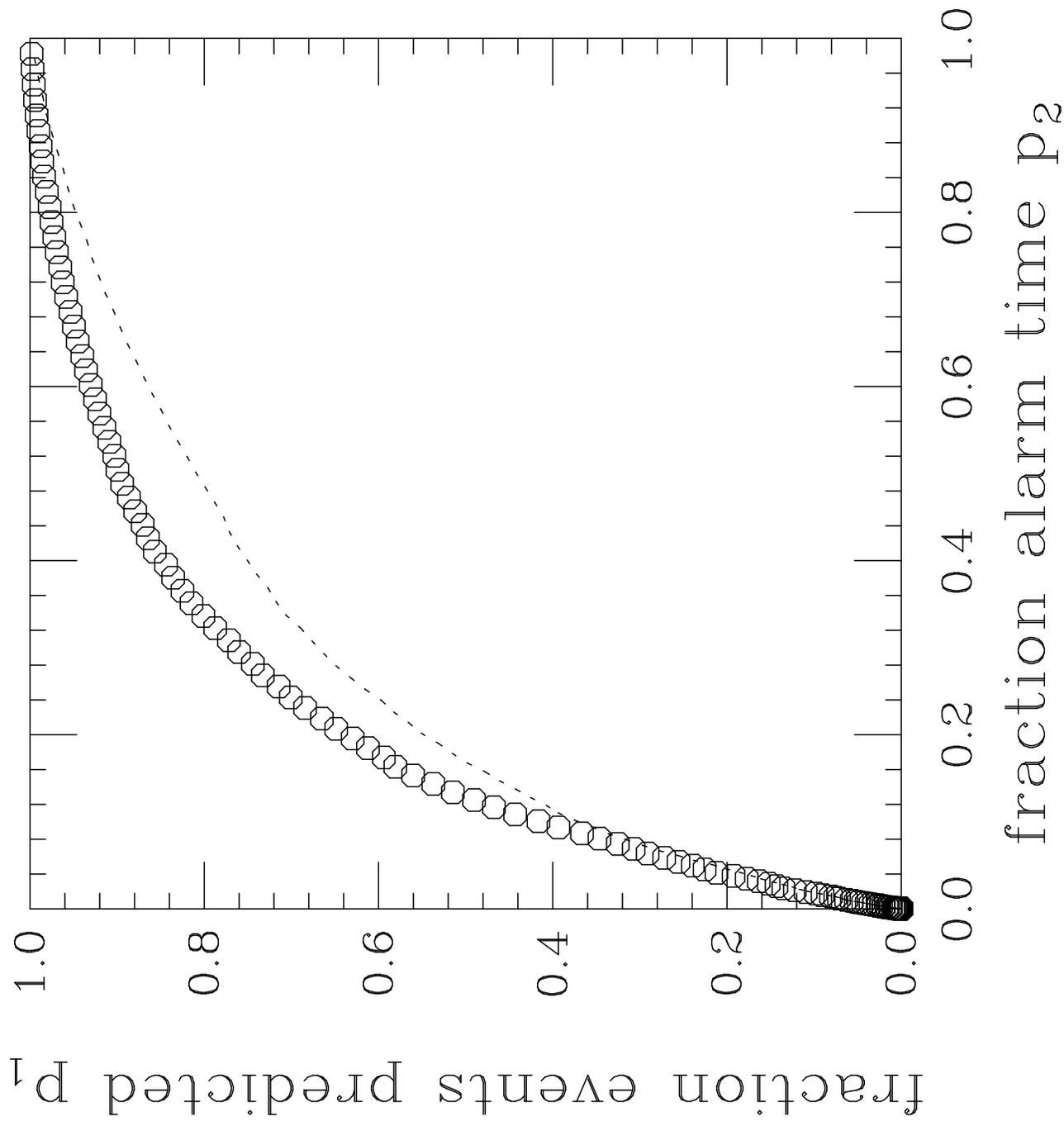
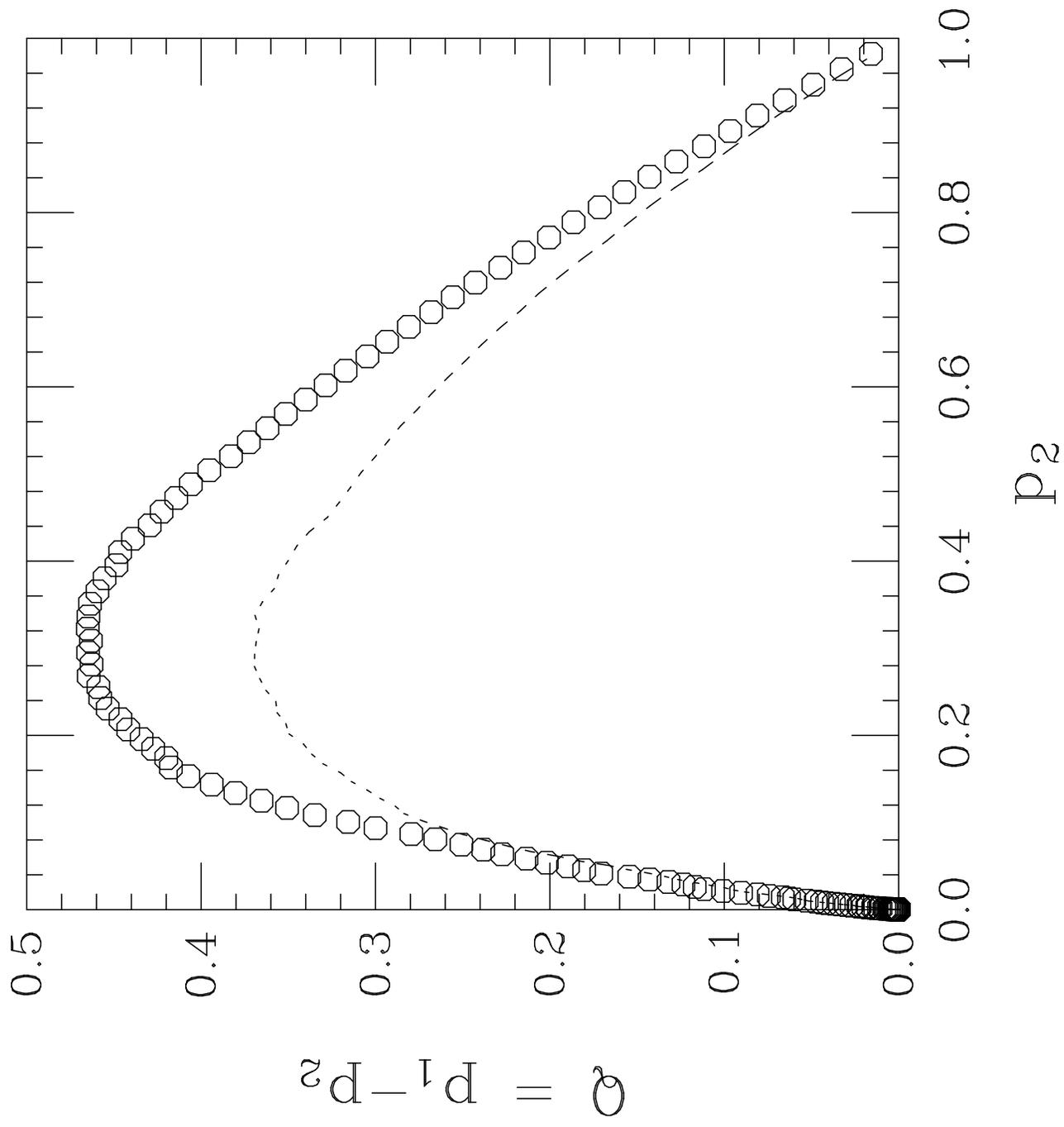
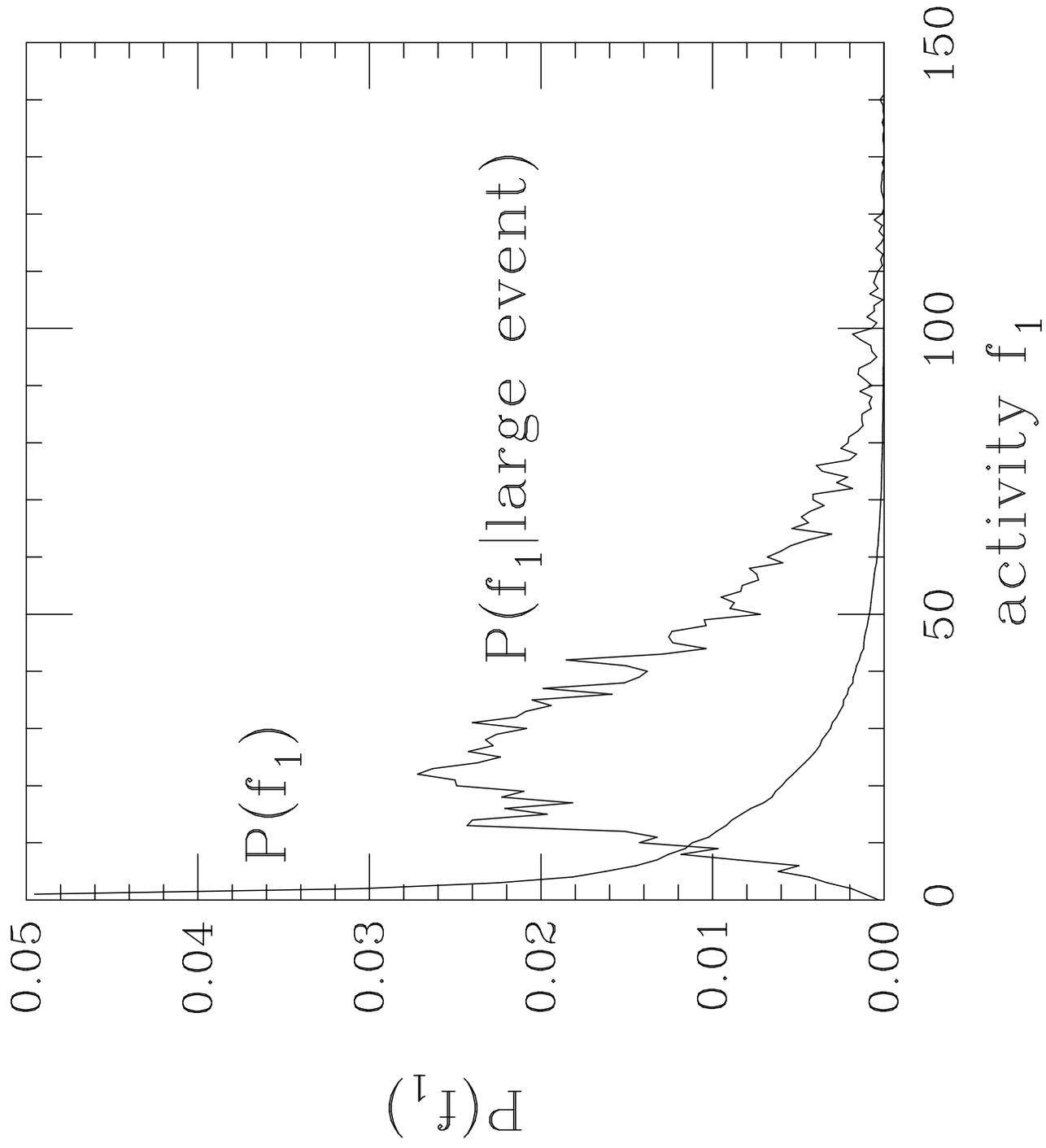
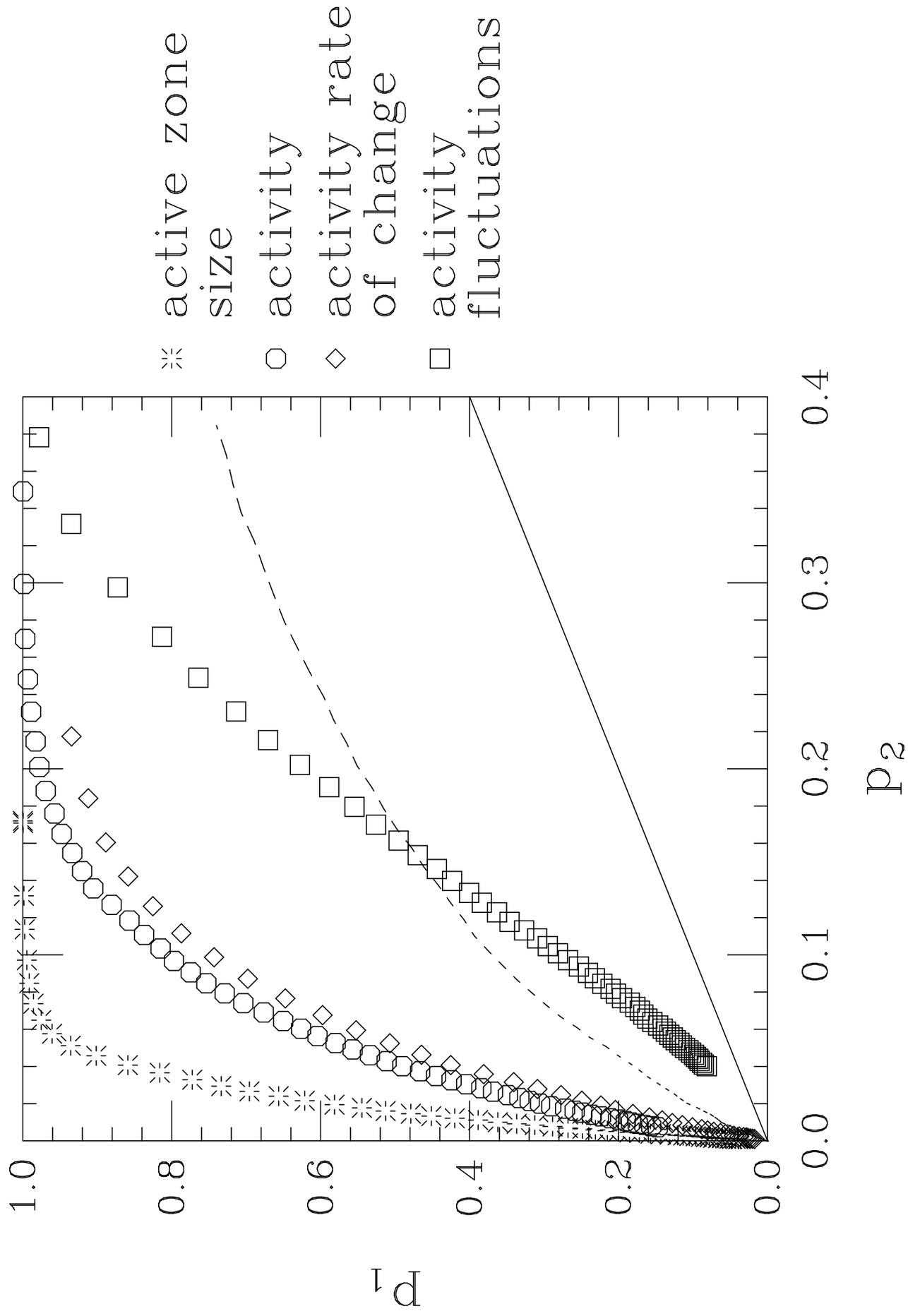
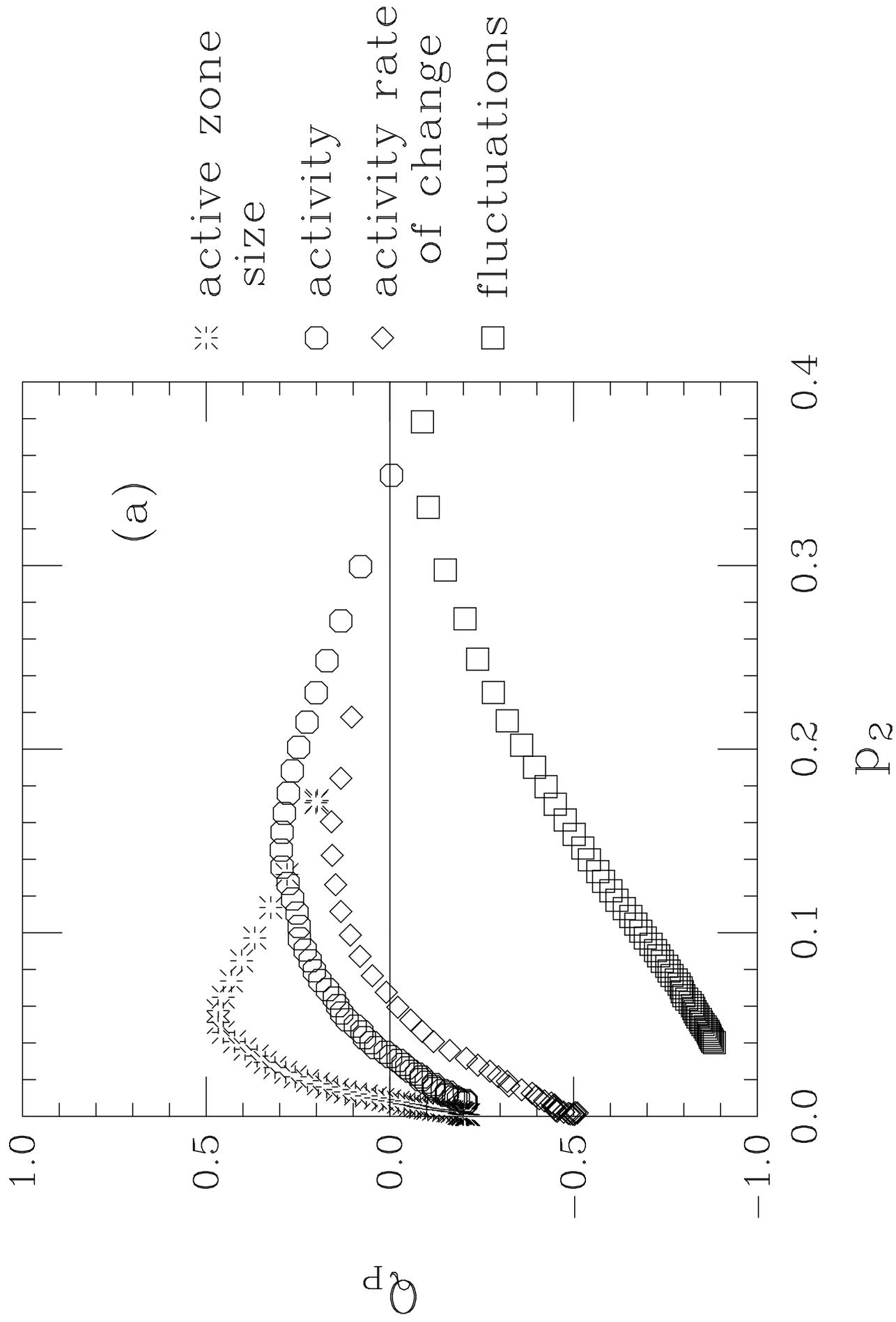
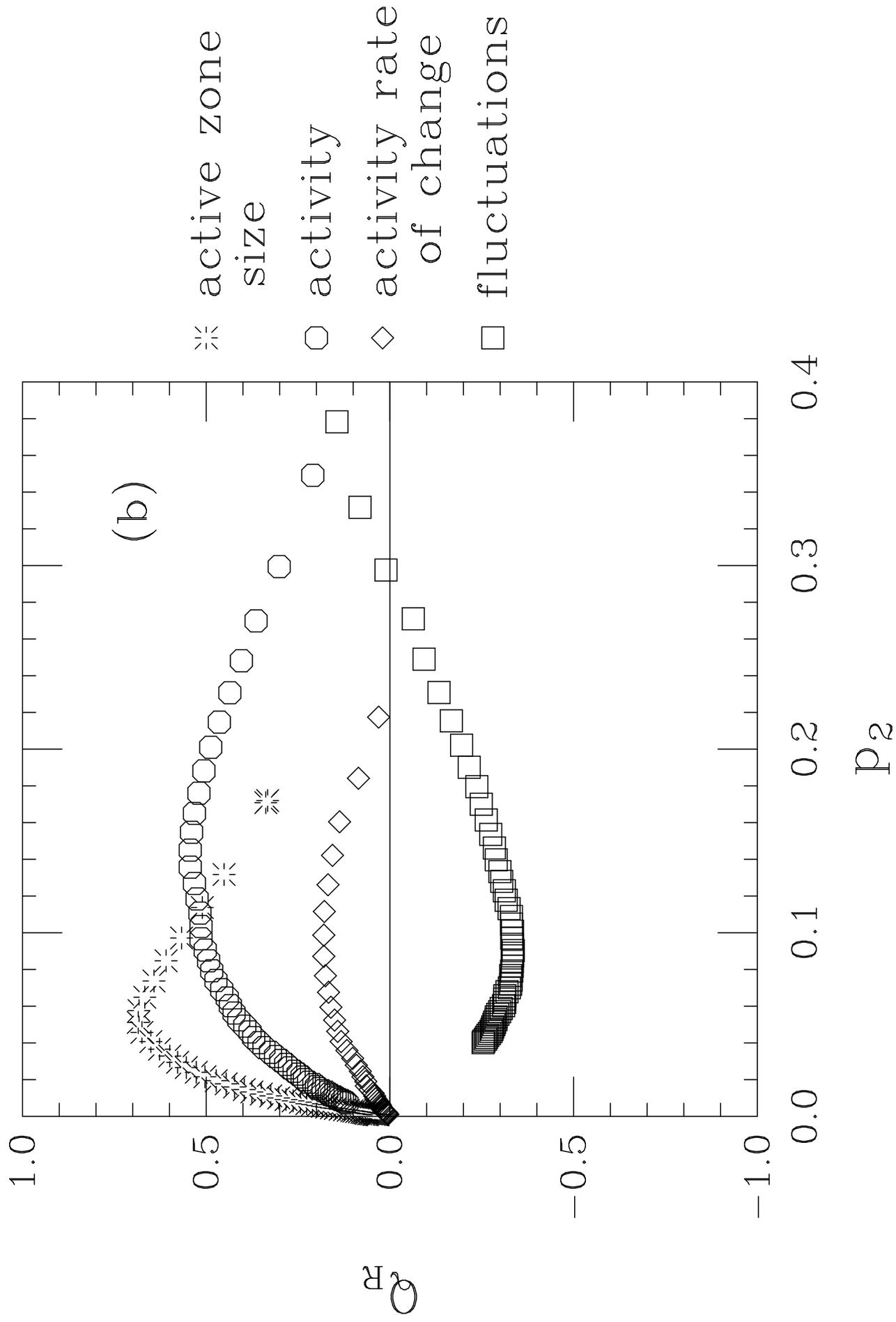
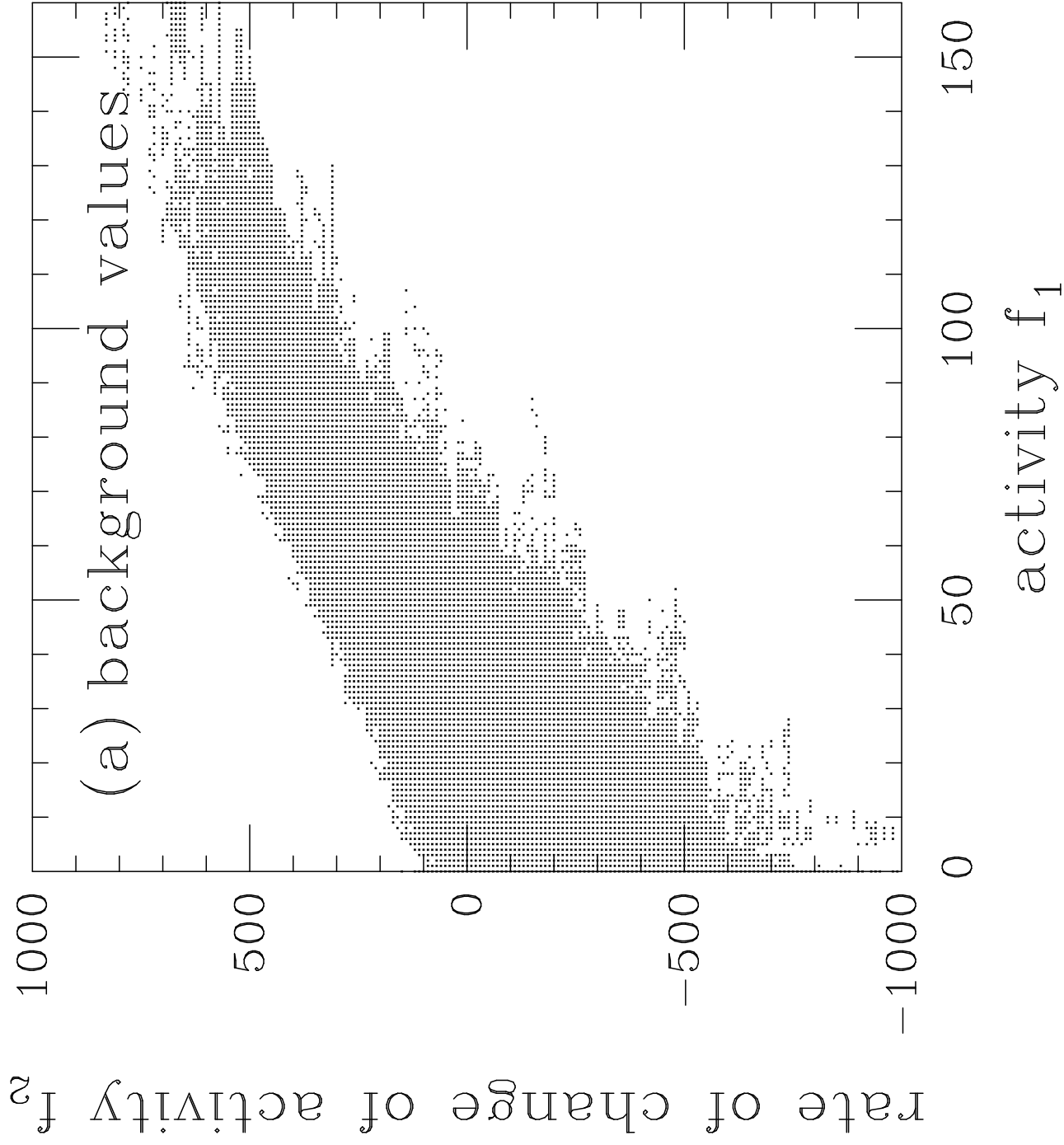
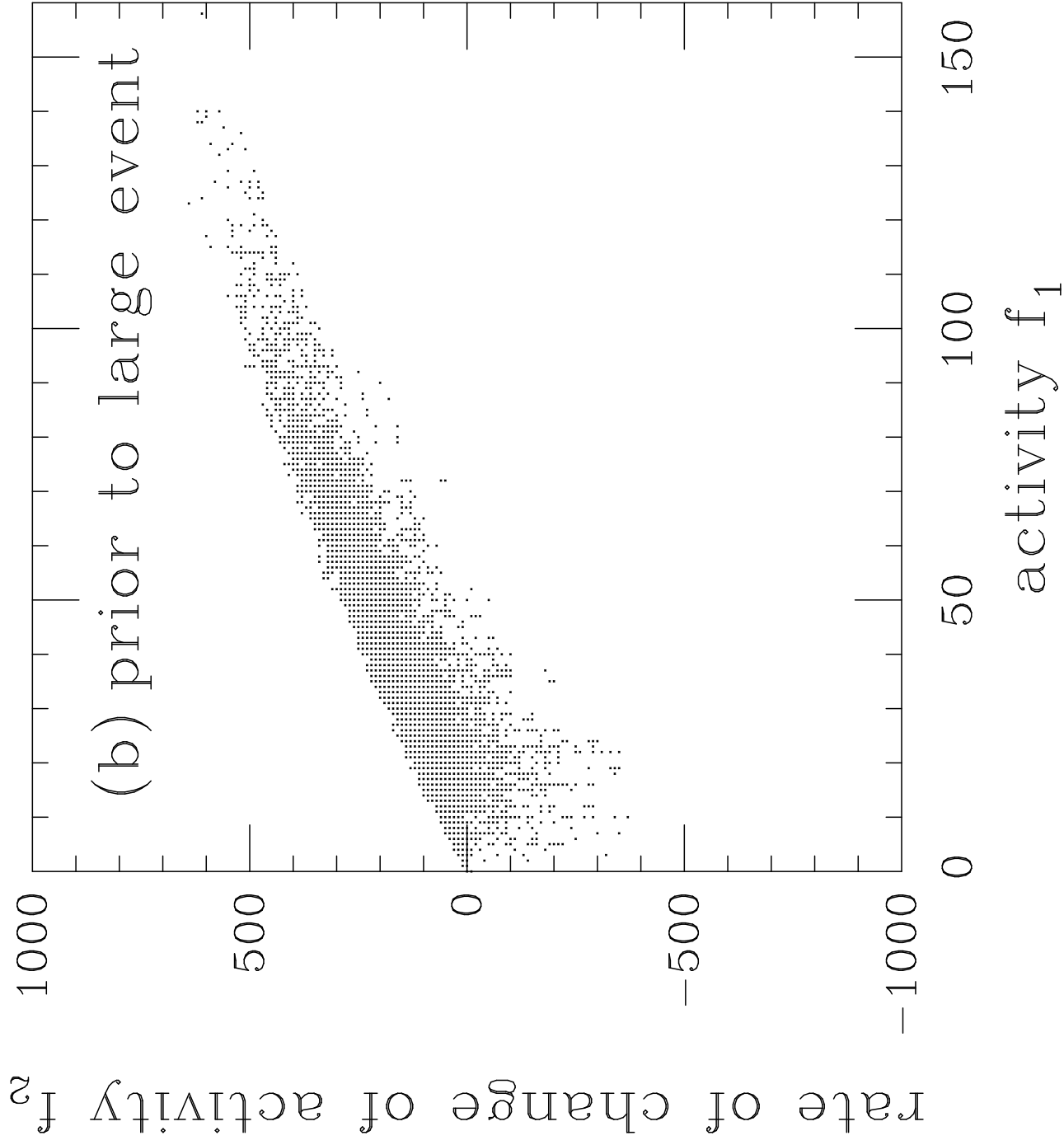
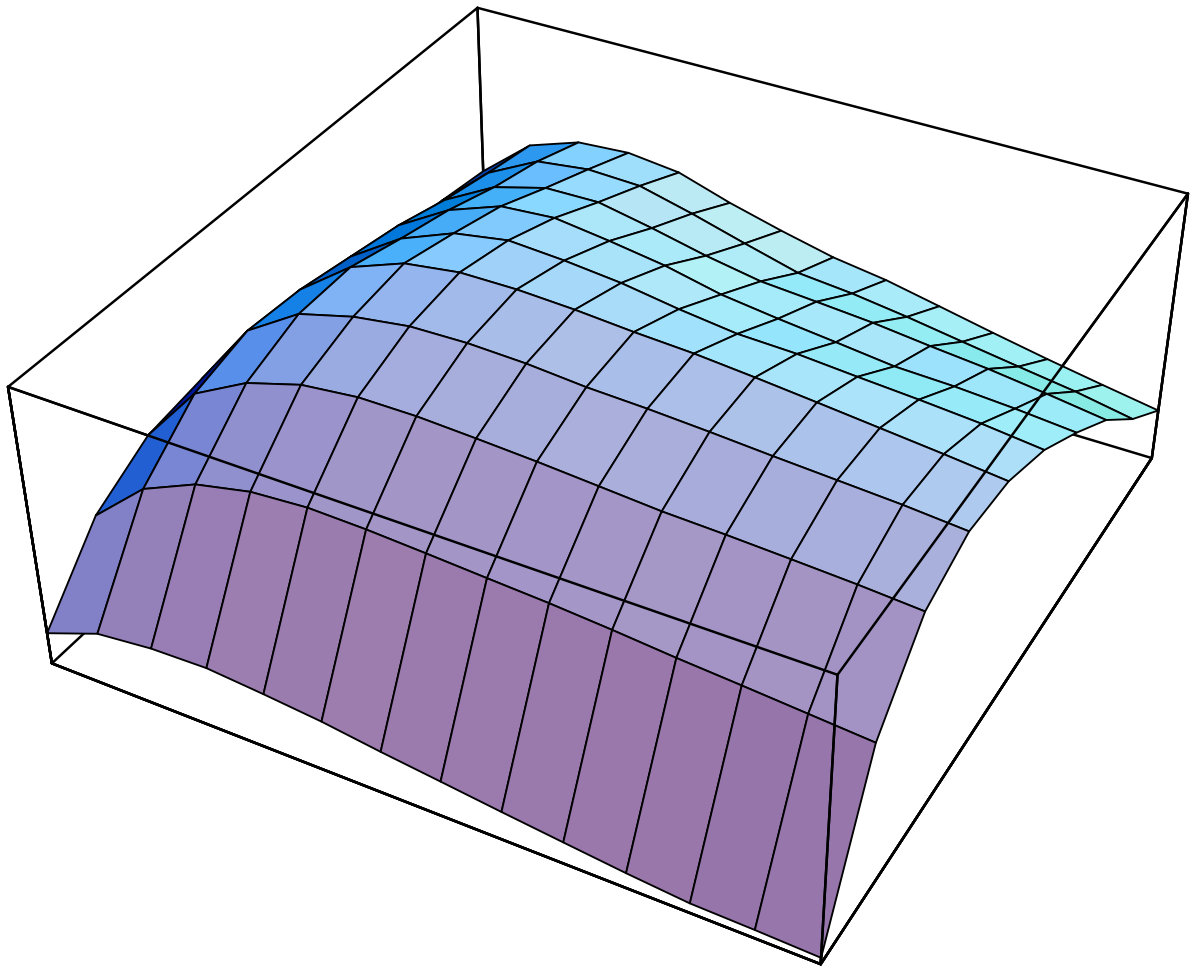
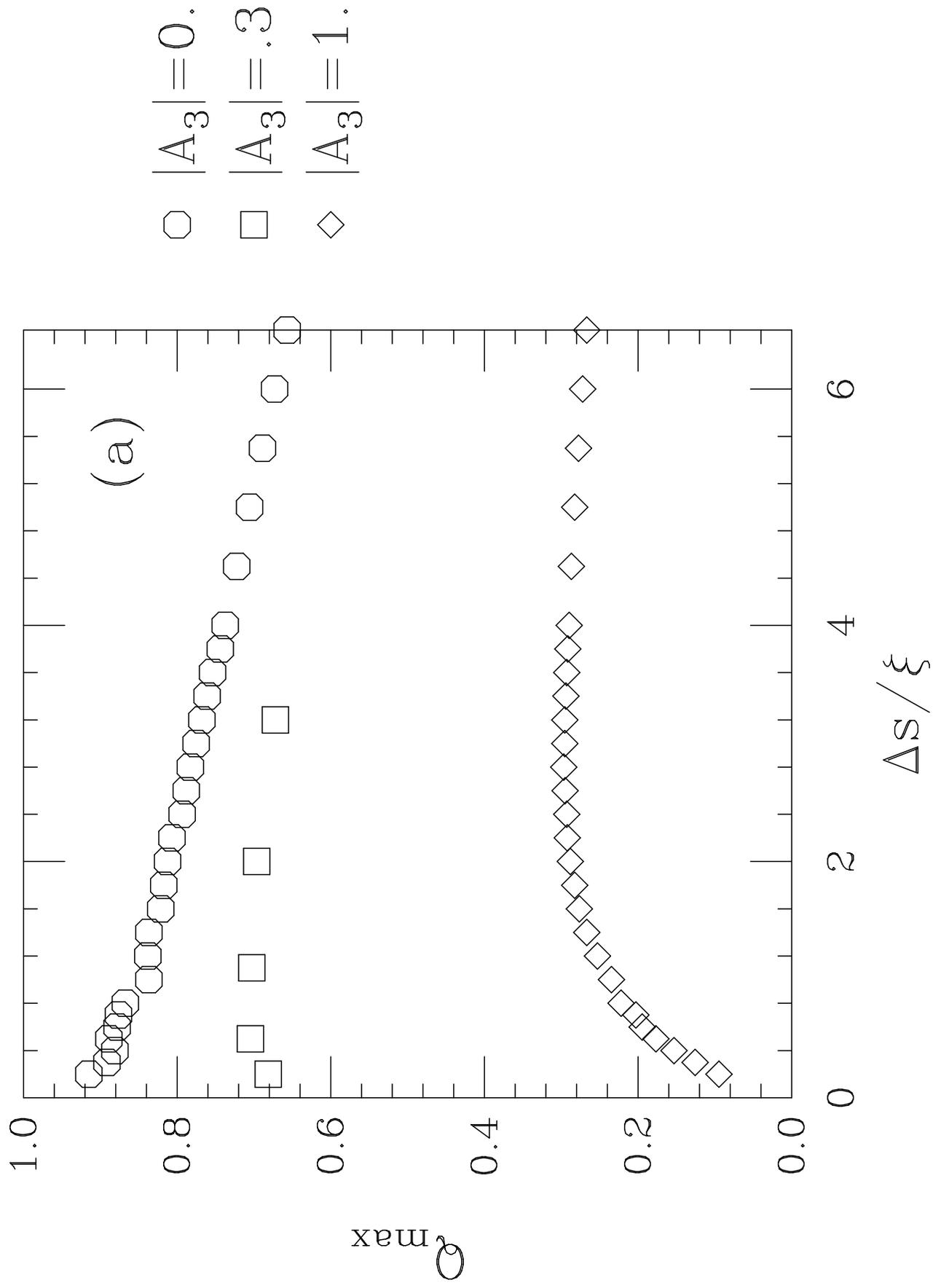
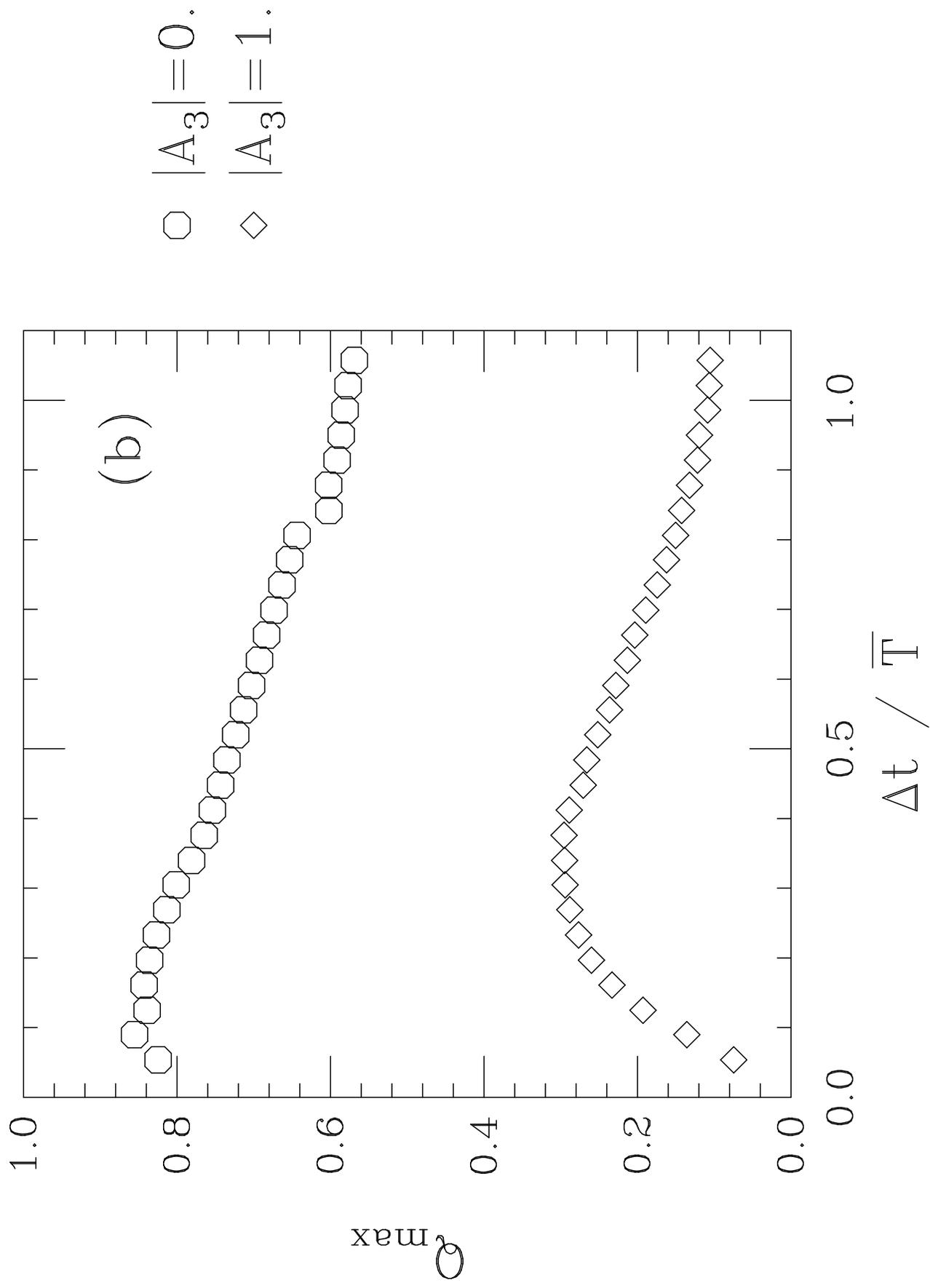
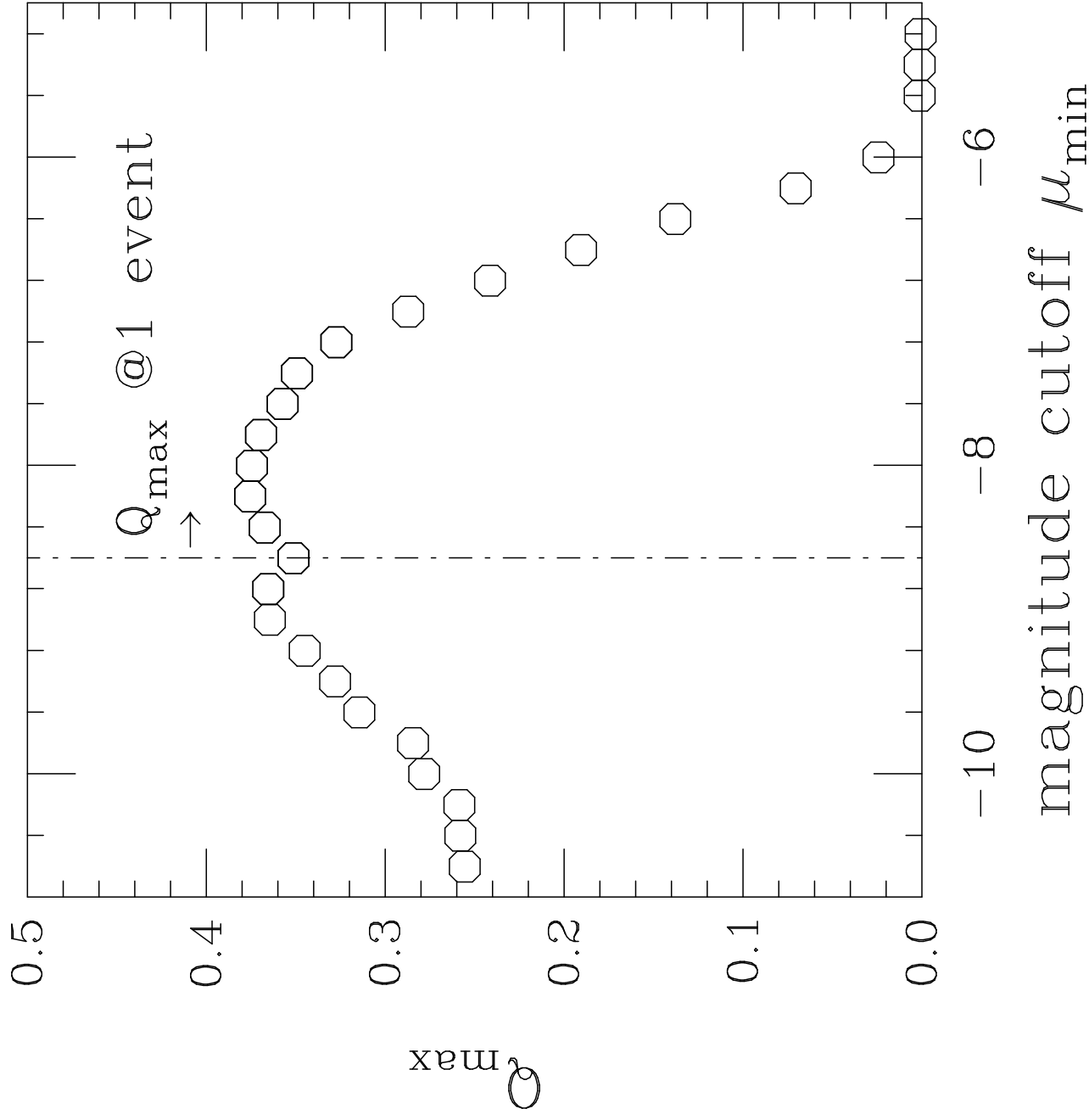
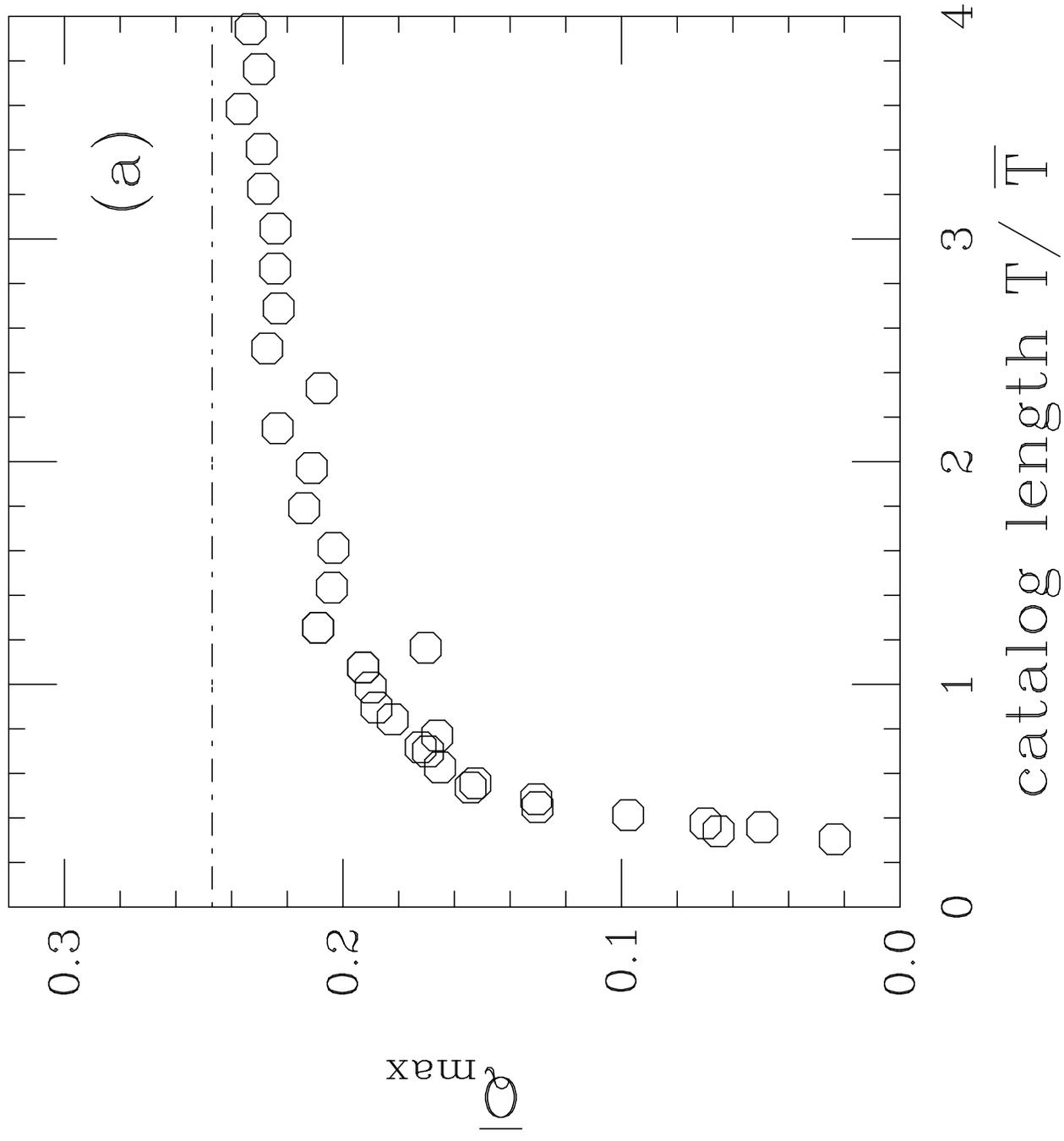
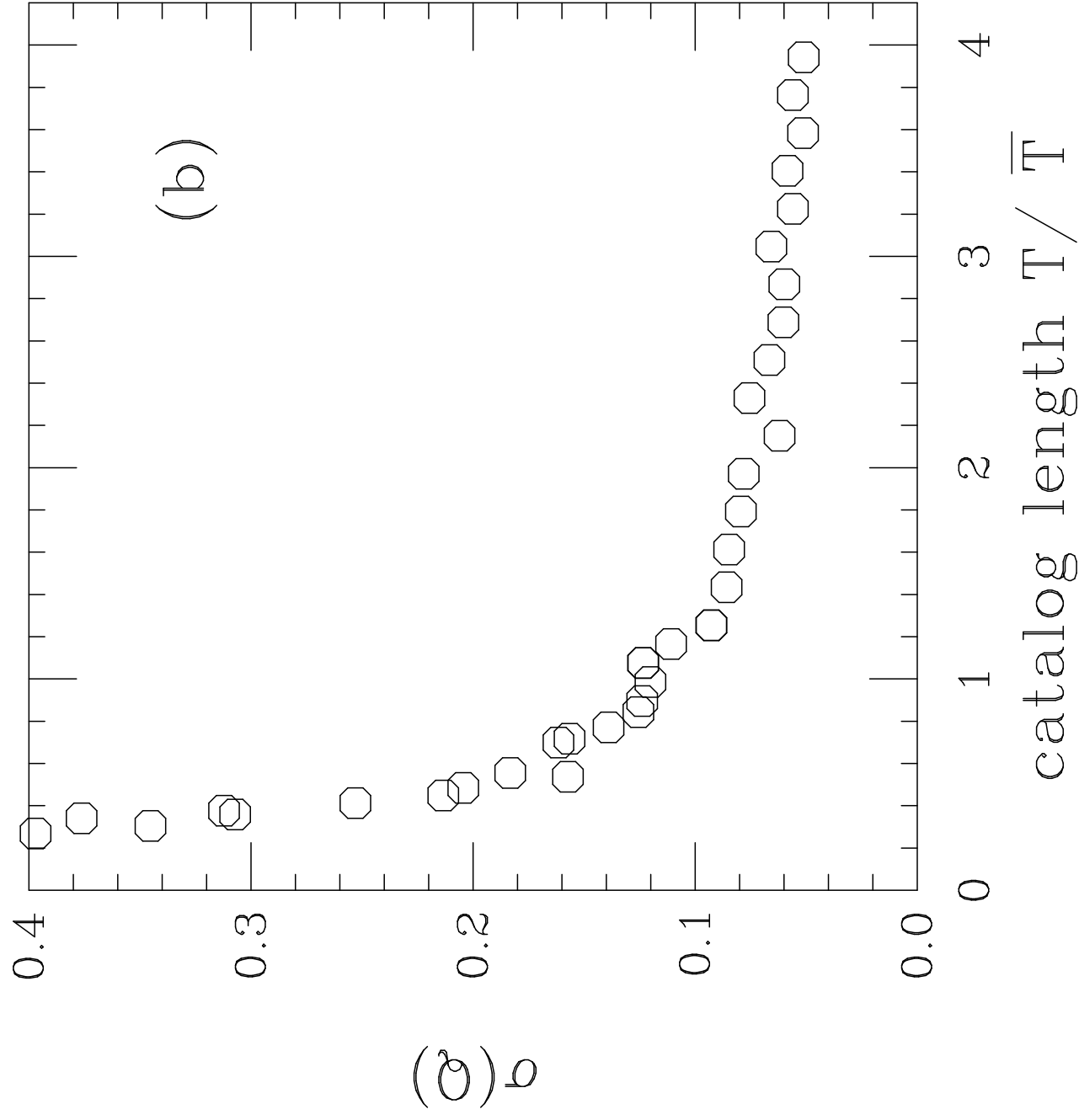
\end